\documentclass[12pt,preprint2,longabstract]{aastex}

\bibpunct{(}{)}{;}{a}{}{,} 

\shorttitle{FEPS: dust processing in TTS systems}
\shortauthors{Bouwman et al.}

\begin{document}

\title{The formation and evolution of planetary systems: \\
Grain growth and chemical processing of dust in T~Tauri systems.}

\author{J.\,Bouwman and Th.\,Henning}
\affil{Max Planck Institute for Astronomy, K\"onigstuhl 17, D-69117, Heidelberg, Germany}
\and
\author{L. A. \,Hillenbrand}
\affil{Department of Astronomy, California Institute of Technology, Pasadena, CA 91125.}
\and
\author{M.R.\,Meyer and I.\,Pascucci}
\affil{Steward Observatory, University of Arizona, 933 N. Cherry Ave., Tucson, AZ 85721-0065}
\and
\author{J.\,Carpenter}
\affil{Department of Astronomy, California Institute of Technology, Pasadena, CA 91125.}
\and
\author{D.\,Hines}
\affil{Space Science Institute,4750 Walnut Street, Suite 205 Boulder, CO 80301.}
\and
\author{J.S.\,Kim and M.D.\,Silverstone}
\affil{Steward Observatory, University of Arizona, 933 N. Cherry Ave., Tucson, AZ 85721-0065}
\and
\author{D.\,Hollenbach}
\affil{NASA Ames Research Center, Moffet Field, CA 94035}
\and
\author{S.\,Wolf}
\affil{Max Planck Institute for Astronomy, K\"onigstuhl 17, D-69117, Heidelberg, Germany}

\begin{abstract}
This paper is one in a series presenting results obtained within the Formation
and Evolution of Planetary Systems (FEPS) Legacy Science Program on the {\it
Spitzer Space Telescope}. Here we present a study of dust processing and growth
in seven protoplanetary disks. 
Our spectra indicate that the circumstellar silicate dust
grains have grown to sizes at least 10 times larger than observed in the interstellar medium, and show evidence for a non-negligible ($\sim$5\% in mass fractions) contribution from crystalline species. These results are similar to those of other studies of protoplanetary disks. In addition, we find a correlation between the strength of the 
amorphous silicate feature and the shape of the spectral energy distribution. 
This latter result is consistent with the growth and subsequent gravitational settling of dust grains towards the disk mid-plane.  
Further, we find a change in the relative abundance of the different crystalline species:
more enstatite relative to forsterite is observed in the inner warm dust
population at $\sim$1 AU, while forsterite dominates in the colder outer
regions at $\sim$5 to 15 AU. This change in the relative abundances
argues for a localized crystallization process rather
than a radial mixing scenario where crystalline silicates are being transported
outwards from a single formation region in the hot inner parts of the disk.
Last, we report the detection of emission from polycyclic aromatic hydrocarbon
molecules in five out of seven sources. We find a tentative PAH band at
8.2~$\mu$m, previously undetected in the spectra of disks around low-mass
pre-main-sequence stars.
\end{abstract}

\keywords{pre-main-sequence stars -- circumstellar disks --- circumstellar dust --- planetary systems}

\clearpage

\section{Introduction}

The circumstellar disks surrounding the pre-main-sequence Brown Dwarfs (BD),
T~Tauri stars (TTS) and Herbig~Ae/Be (HAEBE) systems are believed to be the
sites of ongoing planet formation \citep[e.g.][from here on we will refer to
these disks as protoplanetary disks]{Luhman2007,Natta2007}. As most of the
youngest ($<$1~Myr) solar-mass stars have circumstellar disks \citep{Strom1989},
with typical masses \citep{beckwith1990} and sizes \citep{mcCaughrean1996,
Dutrey1996} comparable to the expected values for the primitive solar nebula,
these disks are the natural candidates for the birth-sites of planets. The
sub-micron sized dust grains present initially in these disks can coagulate to
form larger objects and eventually earth-like planets
\citep[e.g.][]{weidenschilling1997,henning2006}. By deriving the composition of
the circumstellar dust, and identifying the processes governing the chemistry
and coagulation, valuable insights can be gained in the workings of
protoplanetary disks
\citep[e.g.][]{processing,hd100546,Roy2003,Roy2005,Przygodda2003,
Kessler-Silacci2006, Sargent2006} and thus the planet formation process. The
results of these analyses can be compared directly with the investigation of
solar system objects like comets, meteorites and interplanetary dust particles
(IDPs), which preserve a record of the early phases in the evolution of the
solar system.  

With the IRAS and ISO missions, a tremendous advance in our knowledge of
protoplanetary disks was achieved. However, the spectroscopic studies with
these missions were limited to relatively nearby and luminous stars of
spectral type A and B, and provided only limited knowledge of the evolution of
solar-mass systems. With the launch of the  {\it Spitzer Space
Telescope} \citep{Werner2004}, these less luminous systems became accessible to
observations. The FEPS {\it Spitzer} Legacy program probes the circumstellar dust
properties around a representative sample of protoplanetary disks and debris
disks, spanning a wide range of circumstellar disk properties and covering the
major phases of planet system formation and evolution \citep{meyer2006PASP}

We present an analysis of the infrared (IR) spectra of
protoplanetary disks around seven pre-main sequence systems
observed within the FEPS legacy program. These are the only systems among the 
328 FEPS targets showing spectroscopic features from solid state dust components.
\citep[See][for a detailed description of the FEPS parent sample]{meyer2006PASP}.
In Table~\ref{tbl:stars} the astrophysical parameters
of these seven stars are compiled.
Five of the systems,
RX J1842.9-3532,       
RX J1852.3-3700,       
1RXS J132207.2-693812, 
RX J1111.7-7620, and
1RXS J161410.6-230542 ,
were identified in \cite{Silverstone2006} as optically-thick primordial disks on
the basis of their IRAC colors and excess emission at wavelengths shorter than
8~$\mu$m. \cite{Silverstone2006} provide a literature review for the five stars
identified in their study. In addition to these five systems, HD~143006 and
RX~J1612.6-1859A also show evidence for optically-thick disks and solid state
features. We give a literature review for the latter two stars in Appendix~A.
HD~143006 was added to the FEPS sample on the basis of its IRAS excess emission,
in order to search for remnant gas in that system \citep[e.g.][]{pascucci2006}. 
The solid state features in RX~J1612.6-1859A were noticed from visual inspection
of its spectrum.  No other sources in the FEPS sample show
evidence for solid state emission features. Of these 321 sources, most lack
excess in the IRS wavelengths, and a few exhibit optically-thin emission from
debris. No other optically-thick circumstellar disk sources are present in the
data other than the seven with solid state emission features discussed here.

We study the compositional properties of the systems listed in
Table~\ref{tbl:stars} using 5-35~$\mu$m spectra obtained with the Infrared
Spectrograph \citep[IRS;][]{Houck2004} on-board the {\it Spitzer Space
Telescope}. We focus on silicate grain processing (silicates make up the bulk of
the refractory dust mass assuming solar system abundances) and perform a
quantitative analysis of the observed solid state emission.  Using a consistent
analysis method for the entire sample, the derived grain composition of the
individual systems can be compared directly. This enables us look for signs of
silicate grain processing. Apart from analyzing the 10~$\mu$m spectral region,
also accessible to ground-based studies, {\it Spitzer} provides access to the
silicate resonances at longer wavelengths. This enables us to study the dust
composition of not only the warmest dust (temperatures between $\sim$ 500 --
1000 K), located in the inner parts ($\sim0.1-1$~AU) of the protoplanetary disks
around solar-like stars, but also the cold ($\sim$ 100 K) dust component in the
outer regions ($\sim$15~AU) of the disks.  

This paper is organized in the following way: In section~2 we discuss the data
reduction, introduce the main dust components and present the method used to
analyze the Spitzer IRS spectra. In section~3 modelling results are presented
and in section~4 we discuss implications for the evolution of the dust
in the circumstellar disks around solar-mass stars.

\section{Observations and Dust Models}
\subsection{The IRS Low-Resolution Spectra}

We obtained low-resolution ($R=\sim60-120$) spectra with
the IRS instrument on-board {\it Spitzer}. A high accuracy IRS or PCRS
peak-up (with a 1$\sigma$ pointing uncertainty of 0.4" radius) 
was used to acquire targets in
the spectrograph slit, thus minimizing slit losses and assuring high photometric
accuracy to about 10\%. Two nod positions per cycle were obtained in standard staring mode
with a minimum of six cycles per target for redundancy and to allow
the rejection of artifacts introduced by bad pixels or cosmic ray hits. The
integration times were 6s and 14s for the brightest and faintest sources,
respectively. The targets were observed with the full spectral coverage of the
IRS low-resolution instrument between 5.2 and 35~$\mu$m. Beyond 35~$\mu$m the
spectra suffer from excess noise \citep{Houck2004} and can not be used.

Our spectra are based on the {\tt droopres} intermediate data product processed
through the SSC pipeline S12.0.2. Partially based on the {\tt SMART} software
package \citep[][for details on this tool and extraction methods]{Higdon2004},
our data is further processed using spectral extraction tools developed for the
FEPS {\it Spitzer} science legacy program.  As a first step, we correct for the
background emission and stray-light by subtracting the associated pairs of
imaged spectra of the two nodded positions along the slit for each module and
order. In this way we also correct for pixels having an anomalous dark current.
Pixels flagged by the SSC data pipeline as ``bad'' were replaced with a value
interpolated from an 8 pixel perimeter surrounding the errant pixel. We then
extracted the spectra from the resulting set of images using a 6.0 pixel and 5.0
pixel fixed-width aperture in the spatial dimension for the observations with
the short- (5.2 - 14~$\mu$m) and long-wavelength (14-35~$\mu$m) modules,
respectively. The low-level fringing at wavelengths $>$20~$\mu$m was removed
using the {\tt irsfringe} package \citep{lahuis2003}.

Absolute flux calibration was achieved in a self-consistent manner using the
ensemble of FEPS observations. As the majority of the 328 FEPS program stars
exhibit stellar photospheres throughout the shorter wavelength range of the IRS,
this data set is unique for calibrating the IRS instrument. From our full set of
FEPS observations we selected a subset of the spectra which complied with the
following criteria: 1) colors (24-33~$\mu$m) within 1$\sigma$ of an extrapolated
best-fit model photosphere \citep[e.g.][]{meyer2004, carpenter2006} and 2)
Signal-to-noise ratio (SNR) larger than 50 with no artifacts within the spectra.
This resulted for the first order of the short wavelength module (SL1) and the
long wavelength module spectra (LL1, LL2) in a set of 16 stars and for the
shorter wavelength orders (SL2) in a subset of 10 stars. Together with the
stellar photospheric models for these two subsets, we used these spectra to
derive the relative spectral response functions (RSRFs) for the absolute flux
calibration. For the relative (point to point) calibration, we derived RSRFs
using six calibration stars observed at different epochs with the Cohen
photospheric models provided by the SSC, having a superior SNR. This procedure
ensures that the RSRFs used to calibrate our protoplanetary disk spectra have
the highest possible SNR and an estimated absolute flux calibration uncertainty
of $\sim$10\%. For further details on the flux calibration see the explanatory
supplement for the FEPS data, delivered to the {\it Spitzer} data archive (Hines
et al. 2005).

\begin{figure*}
\begin{center}
\resizebox{0.7\hsize}{!}{\includegraphics[angle=0]{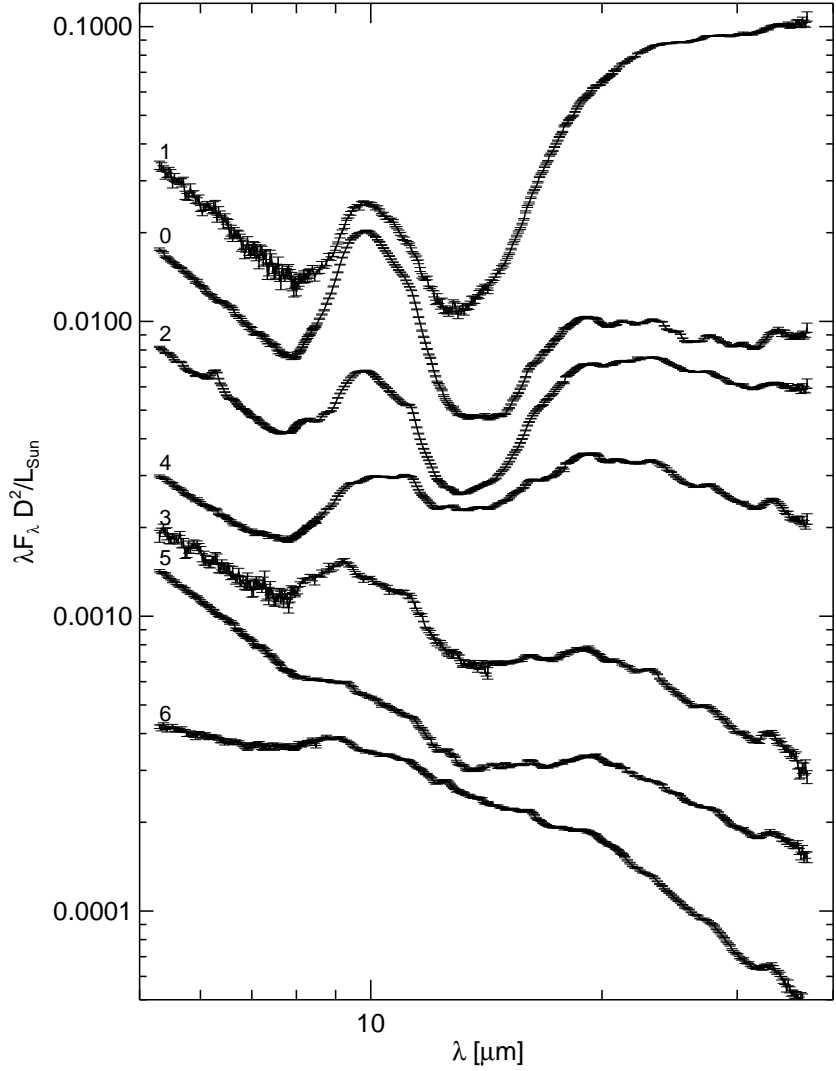}}
  \caption{The spectral energy distributions the the TTS systems observed within
the FEPS legacy program. Shown are the Spitzer low-resolution spectra scaled to
the adopted distance and luminosity of the individual stars as listed in
Table~\ref{tbl:stars}. For clarity the spectra are off-set from each other,
ordered from top to bottom based on the observed slope of the SED, by
multiplying by 20, 5, 0.75, 1, 2.2, 0.25 and 0.2, respectively. }
   \label{fig:specta}
\end{center}
\end{figure*}

\begin{figure*}
\begin{center}
\resizebox{0.7\hsize}{!}{\includegraphics[angle=0]{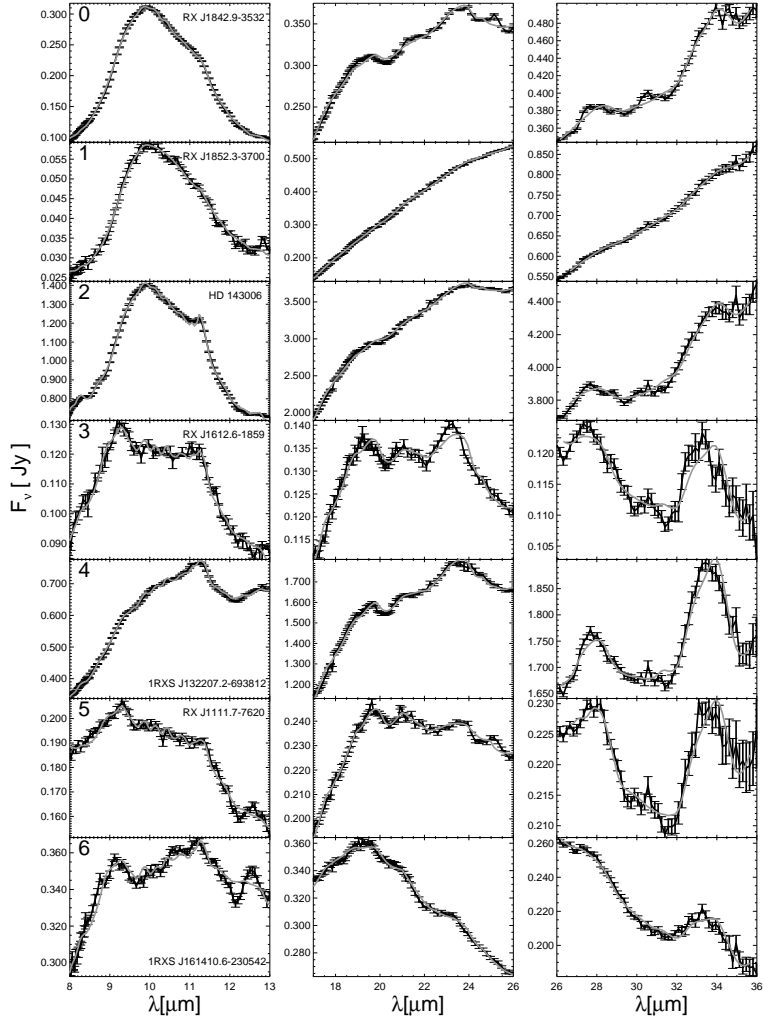}}
  \caption{Spitzer low-resolution spectra  of 7 TTS observed within the FEPS legacy program.
The left panels show the spectra in the wavelength region between 8 and 13~$\mu$m. 
The central panels show the spectral region between 17 and 26~$\mu$m, 
and the right panels the spectral region between 26 and 36~$\mu$m. The spectra are ordered from top to bottom according to the decreasing strength above continuum 
of the 10~$\mu$m silicate band. Also plotted in this figure is a compositional fit to the spectra (light grey lines). For details on the 
model fits see \ref{sec:model}. }
   \label{fig:fit}
\end{center}
\end{figure*}

The resulting calibrated spectral energy distributions are shown in
Fig.~\ref{fig:specta}. A detailed view of these spectra in three wavelength
bands, showing more clearly the solid-state
emission features is presented in Fig.~\ref{fig:fit}. As one can see from
Fig.~\ref{fig:specta}, a wide range in spectral shape and emission feature
strength can be observed. In the next section we will discuss the analysis
methods and the different silicate components and grain models used to interpret
the spectra shown in Fig.~\ref{fig:fit}.

\subsection{Outline of Spectral Analysis Methods}
\label{sec:analysis}

For the analysis and interpretation of our observations we use a two-fold
approach. First, we characterize the spectra by measuring the
strengths and positions of spectral features. Second, we decompose the observed emission features to  determine the relative contribution of different silicate grains.
For the spectral characterization we employ a method similar to schemes previously used
to analyze IR spectra from the Infrared Space Observatory
\citep[e.g.][]{molster2002B,molster2002A} and ground-based observations
\citep[e.g.][]{Roy2005}. To measure the emission band strengths and positions,
we first fit a low order polynomial (second order in the 10~$\mu$ region, fifth order for $\lambda > 15$~$\mu$m) to the spectral data points excluding obvious
spectral features (e.g. the continuum). As a next step we normalize our spectra to the fitted continuum as 
$F_{\nu,\mathrm{norm}} = 1 + (F_{\nu,\mathrm{obs}}-F_{\nu,\mathrm{cont}})/<F_{\nu,\mathrm{cont}}>$,
where $F_{\nu,\mathrm{obs}}$ is the observed 
Spitzer flux,  $F_{\nu,\mathrm{cont}}$ is the fitted continuum, and $<F_{\nu,\mathrm{cont}}>$ the mean continuum value over the fitted wavelength interval. This normalization allows us to directly compare the different spectra and, ensures 
that the shape of the spectral features remains identical to that in the original, 
unnormalized, spectra.

To characterize the 10~$\mu$m amorphous silicate
band we determined the fluxes at different wavelengths across the silicate
feature in the normalized spectra. The results of this analysis are presented in Section~3.  Amorphous
silicates also show a band at around 18~$\mu$m. However, this band blends with
the (often rising) continuum emission, as can for instance be seen in
Fig.~\ref{fig:specta}. This makes it difficult, if not impossible, to determine the
band strength and position with sufficient accuracy for a meaningful analysis.
We, therefore, opted not to use this band in our analysis.  At longer
($>$20~$\mu$m) wavelengths the emission features are mainly produced by
crystalline silicates.  To measure the observed band positions and strengths, we
simultaneously fitted a Gaussian to each of the main spectral features in the normalized
spectra. The results of this analysis are also presented in Section~3.  For the
crystalline silicate emission bands in the 10~$\mu$m wavelength region, a
slightly modified analysis method has to be applied. In this wavelength range,
the emission bands overlap with, and can be dominated by emission from small
amorphous silicate grains. Also, PAH molecules may contribute to the observed
emission in the short wavelength range, overlapping with the emission features
of the crystalline silicates. Therefore, the crystalline contribution can only
be estimated by fitting a detailed grain model (discussed in the next section)
determining the contribution of the amorphous silicates and PAH molecules. After
subtracting the fitted contribution of the amorphous silicate and PAH molecules,
we applied an analysis for estimating the different crystalline silicate band
strengths in the 10~$\mu$m wavelength region, similar to that for the longer
wavelength crystalline bands. 

\subsection{Dust Models}
\label{sec:model}

The grain model and analysis method for determining the physical properties of
the dust grains contributing to the observed emission are similar to those
successfully applied in previous studies of the 10~$\mu$m region \citep[see][for further
discussion]{processing,Roy2005}. Here,  we also apply this
method to the silicate emission at longer ($\sim$20-30~$\mu$m) wavelengths. Dust in 
protoplanetary disks has most likely the structure of highly irregular
aggregates containing many different dust constituents, like
the interplanetary dust particles collected in the Earth's upper atmosphere. Calculating the optical properties of such
complex structure has proven to be extremely difficult \citep[ e.g.][]{henning1996}. 
However, assuming that the aggregates are extremely porous, the
individual constituents making up the aggregate may interact with the radiation
field as separate entities, as is the case of IDPs \citep[][]{molster2003}. 
Therefore, we assume that the observed emission can be represented by the sum of the
emission of individual dust species. Table~\ref{tbl-species} summarizes the dust
species used in our analysis, including all dust species commonly identified in protoplanetary disks \citep[e.g.][]{processing, Roy2005, Sargent2006}.

Crystalline silicates such as forsterite (Mg$_2$SiO$_4$) and enstatite
(MgSiO$_3$) have many strong and narrow resonances in the wavelength range
covered by the IRS spectrograph. These rotational/vibrational bands of the crystalline dust species allow for an much more accurate determination of the chemical composition, grain size and shape/porosity compared to the amorphous silicates. 
We find, like in the above mentioned studies, that the pure magnesium end-members of the olivine and pyroxene families give the best match to the observed spectral features. 
Based on the comparison between the in our spectra observed band positions and strengths, and laboratory measurements \citep{Fabian2001,Koike2003}, we find no evidence for iron containing crystalline silicates. Still a few open questions remain. Enstatite comes in the form of clino- and ortho-enstatite, both having a different crystalline structure. However, both forms show very similar emission bands at around 10~$\mu$m, making it hard to distinguish between them \citep[e.g.][]{jaeger1998,koike2000}. Only at the longer wavelengths a clear difference between the emission properties can be observed. Unfortunately, as we will discuss in the next chapter, no clear enstatite bands can be observed at the longer wavelengths, making it difficult to determine the exact enstatite structure. We, therefore, have chosen to use clino-enstatite, allowing for a direct comparison between our results and the study by \cite{Roy2005} of a sample of Herbig~Ae/Be stars. A similar problem occurs for silica dust grains.
Silica has nine polymorphs among which are quartz, tridimite and cristobalite, differing in crystal structure.  While of these silicates quartz is the most common form found on Earth, the common form found in interplanetary dust particles (IDPs) is tridimite \citep{rietmeijer1988} and most likely also the most common form in protoplanetary disks.
To our knowledge, no similar quality laboratory measurements of tridimite exists, enabling us to perform an identical analysis as for the other dust species.
However, if one compares the amorphous form of silica and quartz, we observes similar bands. It might well be, that like in the case of enstatite, it will be difficult to tell the different forms of silica apart. As quartz is unlikely to exist in protoplanetary disks we have opted to use an amorphous silicate with the stoichiometry of silica.

As already was noted in previous studies \citep[e.g.][]{processing}, the observed emission bands of forsterite, enstatite and silica can not be reproduced assuming 
simple homogeneous spherical grains, but that one has to adopt an inhomogeneous grain
structure and/or non-spherical grain shape. Here we use a distribution of hollow spheres
\citep[DHS;][]{min2005} which gives an excellent match to the observed band
positions and shapes. DHS has the advantage over the widely used continuous distribution of ellipsoids \citep[CDE;][]{bohren1983} that it is also defined outside of the Rayleigh limit, making it possible to investigate the effects of grain size on the emission properties of the grains. One should realize that the above discussed grain models such as DHS or CDE should not be taken as an exact representation of the structure/shape of a grain, but rather as a statistical description of the grain shape and structure and its deviation from sphericity and homogeneity; \citep[see][ for a further discussion of grain shape and porosity effects]{voshchi2006,min2006}. 

For the amorphous iron/magnesium silicates the exact composition and grain shape
can not be constrained as well as for the crystalline silicates. Models of the
broad amorphous silicate resonances at $\sim$10~$\mu$m and $\sim$18~$\mu$m are
to some extent degenerate. Silicate models with varying magnesium to iron ratios
and different grain shapes/homogeneity can reproduce the observed bands
\citep[e.g.][]{Min_ISM}. To allow for a direct comparison with previous studies,
we assume here that the amorphous silicates have an equal magnesium to iron
atomic ratio, and stoichiometries consistent with olivine and pyroxene. Further,
for the amorphous iron-magnesium silicates we assume homogeneous, spherical
grains for which we calculate the absorption coefficients using the Mie theory
\citep[e.g.][]{bohren1983} and the optical constants listed in
Table~\ref{tbl-species}. 

\begin{figure}[t!]
\begin{center}
\resizebox{0.8\hsize}{!}{\includegraphics[angle=0]{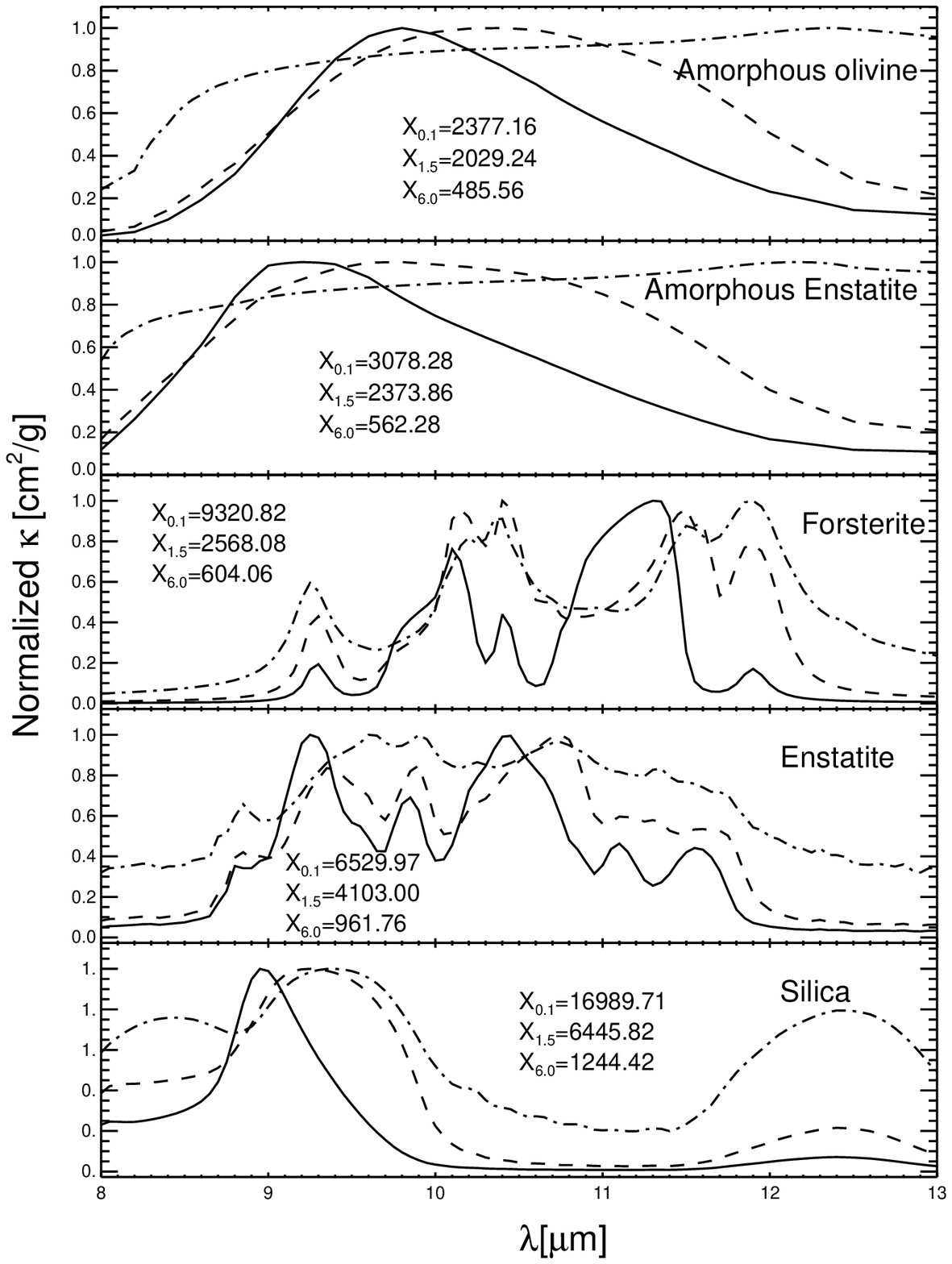}}
\caption{To unity scaled opacities  for the different dust species 
  listed in Table~\ref{tbl-species}. Figure 3a shows the opacities between 
8 and 13~$\mu$m, Figure 3b between 17 and 26~$\mu$m, and Figure 3c between 26 and 36~$\mu$m. In each panel the solid lines represent the opacities for a grain size of 0.1~$\mu$m, the dashed lines the opcities for a 1.5~$\mu$m grain size, and the dot-dash lines the opcities of 6~$\mu$m sized grains. Also indicated in each panel are the scaling factors for each of the opacity curves, used in the normalization to unity.}
\label{fig:kappa_short}
\end{center}
\end{figure}

To take into account the effect of grain size, we calculated for each of the
different grain species the opacities for three volume equivalent grain radii of
0.1~$\mu$m, 1.5~$\mu$m and 6.0~$\mu$m, respectively. These grain sizes sample
the range of spectroscopically identifiable grain sizes at infrared wavelengths
at the SNR of our data. The resulting calculated opacities for three successive
wavelength regimes are shown in Figs.~\ref{fig:kappa_short}a,
\ref{fig:kappa_short}a and \ref{fig:kappa_short}b. We present our analysis of
the various dust species in Sections ~\ref{sec:coagulation}, \ref{sec:settling},
and ~\ref{sec:crystalline}. 

Apart from the silicate dust species, PAH molecules also
contribute to the observed emission bands within the spectral range of the IRS
instrument. These large molecules are not in thermal
equilibrium with the radiation field, but are stochastically heated. Though many
of the observed spectral features can be linked to specific vibrational modes
and ionization states of the molecules, the exact composition remains unclear.
As our main focus in this paper is the solid-state component in the protoplanetary disks,
we opted to use a template spectrum, based on observed profiles by
\cite{peeters2002} and \cite{diedenhoven2004}, for estimating the contribution of PAH molecules
to the spectra. We present this analysis in Section~\ref{sec:pah}.

The thermal infrared emission from protoplanetary disks is
determined by the complex spatial distribution of individual dust species,
and the radial and vertical optical depth and temperature distributions. A
completely self-consistent model of protoplanetary disks, such as a full
radiative transfer calculation, would also have to account for processes such as
radial and vertical mixing of circumstellar material and chemical and structural
alteration of the dust grains under varying physical conditions in different
parts of the disk. Further, to uniquely constrain such models, spatially
resolved observations as a function of wavelength \citep[e.g.][]{Thamm1994} are necessary.  
Instead, we employ a more simplistic approach which has been demonstrated to
successfully reproduce the silicate emission features of HAEBE and TTS systems.
Detailed self-consistent models \citep[e.g.][]{menshchikov1997,CG,dullemond2001}
have shown that the emission features originate from a warm optically thin disk
surface layer on top an optically thick disk interior, producing continuum
emission. The basic assumption of our simple model is that the disk emission
features can be reproduced by a sum of emissivities at a given temperature
T$_\mathrm{dust}$, and the continuum emission by the emission of a black body at
a given temperature T$_\mathrm{cont}$, which could be interpreted as the typical
color temperature of the underlying continuum at a given wavelength. 

We thus fit the following emission model 
using a linear least square minimization to the Spitzer low-resolution spectra:
\begin{eqnarray}
\label{eq:1}
F_\nu &=& B_\nu(T_\mathrm{cont}) C_0  \nonumber \\
&+& B_\nu(T_\mathrm{dust}) \left(\sum_{i=1}^3\sum_{j=1}^5 C_{i,j} \kappa_\nu^{i,j} \right) \nonumber \\
&+& C_\mathrm{PAH} I_\nu^\mathrm{PAH},
\end{eqnarray}
where $B_\nu(T_\mathrm{cont})$ denotes the Planck function at the characteristic
continuum temperature $T_\mathrm{cont}$, $B_\nu(T_\mathrm{dust})$ the Planck
function at the characteristic silicate grain temperature $T_\mathrm{dust}$,
$\kappa_\nu^{i,j}$ is the mass absorption coefficient for silicate species $j$
(five in total) and grain size $i$ (three in total), $I_\nu^\mathrm{PAH}$ is the
PAH template emission spectrum, and $C_0$, $C_{i,j}$ and $C_\mathrm{PAH}$ are
the weighting factors of the continuum, the silicate components and the PAH
contribution, respectively. For the single temperature approximation to be
meaningful, the width of the spectral region, and thus the temperature range,
needs to be sufficiently narrow. Therefore, we have split the observed spectra
into three bands: from 6 to 13~$\mu$m, from 17 to 26~$\mu$m, and from 26 to
36~$\mu$m. Apart from the previous considerations, these wavelength boundaries
also ensure that both continuum and spectral features can be well characterized,
and that multiple bands of the main crystalline silicates can be modeled
simultaneously. This latter point is important as it reduces the degeneracy problem of overlapping emission bands of the different dust species. 
We have limited the maximum temperature in our fits to 1500~K,
above which all silicate dust is expected to be vaporized.

Fig.~\ref{fig:fit} shows the resulting best-fit models for the three spectral
regions. As one can see, a good agreement between model and observations can be
obtained using the dust species discussed in this section, without the need for
additional components. The resulting model parameters are listed in
Table~\ref{tab:t2_fs:fits}. We estimate the uncertainties in the derived band
strengths and positions and the best-fit dust compositions by a monte-carlo
method as discussed in Section 15.6 of\citep[][]{NumRec}: We generate a set of
100 spectra for each observation  by randomly adding Gaussian noise with a
1~$\sigma$ distribution equal to the error in the spectral extraction determined
for each spectral data point. For each of these simulated spectra, we applied
the analysis methods described above. Our quoted results represent the mean
values of the derived quantities and the error is the 1~$\sigma$ standard
deviation in the mean. 

\section{Results}

Following the dust modelling and spectral decomposition, our main science
results are in the areas of dust coagulation from the analysis of the 10 $\mu$m
amorphous silicate feature, dust settling from the additional analysis of the
overall SED shape, crystallinity from the analysis of the various silicate dust
species, and PAHs, which remain in the residual emission after subtracting the identified
silicates.

\subsection{Coagulation in protoplanetary disks}
\label{sec:coagulation}

Grain size strongly influences the infrared opacities, as can be seen in
Fig.~\ref{fig:kappa_short}a to \ref{fig:kappa_short}c. Typically, for the
observations presented here, grain sizes (assuming compact grains) of up to
10~$\mu$m can be inferred spectroscopically. This is two orders of magnitude
larger than the grain sizes typically derived for dust grains in the ISM ($\le 0.1$~$\mu$m). For larger grains,
the spectral signatures become too weak to distinguish them from continuum
emission. Amorphous iron-magnesium silicates, making up the bulk of the solid
state material in the ISM, show two broad spectral features at about 10~$\mu$m and
18~$\mu$m from which grain sizes can in principle be determined. The latter
band, however, often blends with continuum emission from the disk interior and
colder dust grains at larger radii, making it difficult to determine the
grain size. For this reason, the 10~$\mu$m silicate band gives
the most accurate information concerning amorphous silicate grain sizes. Many
previous studies have shown that both the shape \citep[][]{processing} and
strength \citep[][]{Roy2003,Roy2005,Przygodda2003,schegerer2006,Kessler-Silacci2005,
Kessler-Silacci2006} of the 10~$\mu$m silicate features observed in HAEBE and TTS 
systems are mainly determined by the size of the amorphous silicate grains.

\begin{figure*}[t]
\begin{center}
\resizebox{0.9\hsize}{!}{\includegraphics[angle=0]{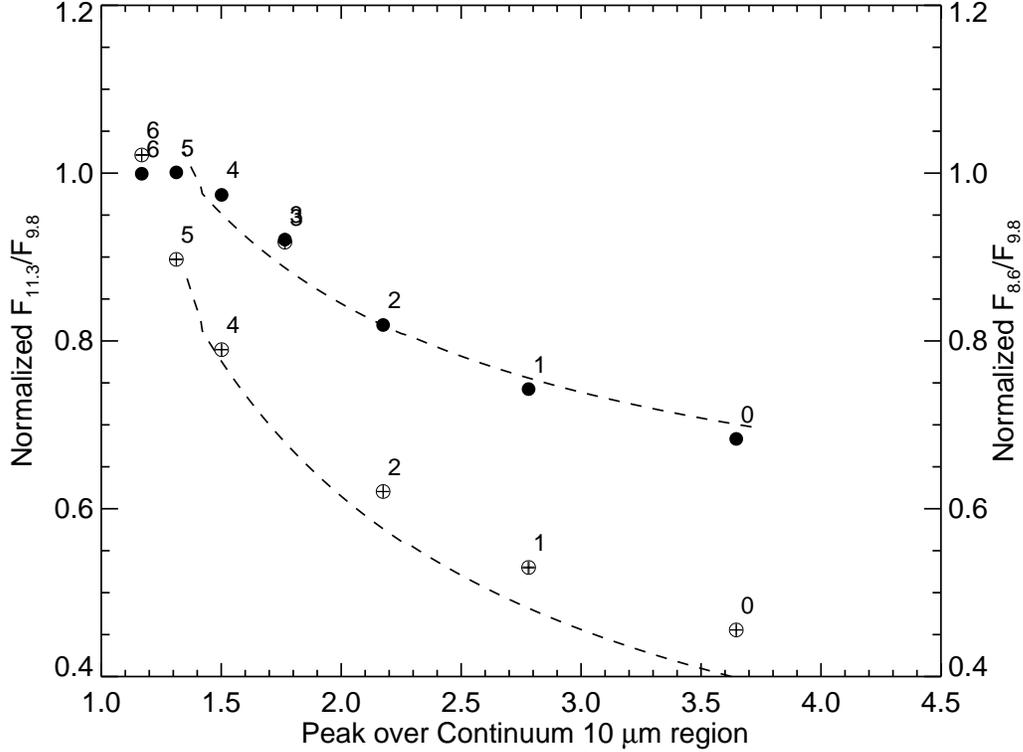}}
  \caption{Correlations between the strength and the shape of the 10~$\mu$m
silicate band. The filled symbols show the correlation between the peak over
continuum ratio of the 10~$\mu$m silicate band and the ratio of the normalized
flux at 11.3~$\mu$m over 9.8~$\mu$m (left axis). The open symbols show the
correlation between the peak over continuum ratio and the ratio of the
normalized flux at 8.6~$\mu$m over 9.8~$\mu$m (right axis). Note that the formal error on the plotted quantities is smaller then the size of the used symbols. The numbers
correspond to the ID numbers of our target stars as listed in Table~1. Also
plotted in this figure, represented by the dashed lines, is the calculated
behavior for both flux ratios of amorphous olivine grains for a continuous
changing grain size from 0.1~$\mu$m (lower right) to 2.0~$\mu$m (upper left).}
   \label{fig:cor_shape}
\end{center}
\end{figure*}

Fig.~\ref{fig:cor_shape} shows the correlation between the shape and strength of
the 10~$\mu$m silicate feature as measured by the relation between the peak over
continuum strength of the 10~$\mu$m silicate feature and the ratio of normalized
fluxes at 11.3 to 9.8~$\mu$m (filled symbols) and at 8.6 to 9.8~$\mu$m (open
symbols). We performed a Kendall's $\tau$ test on the data plotted in Fig.~\ref{fig:cor_shape}. This test computes the probability that (x,y) pairs of data
are correlated in the sense of the relative rank ordering of the x-values
compared to the relative rank ordering of the y-values \citep{NumRec}.  A value of
Kendall's $\tau$ was computed along with the two-sided probability
that the variables are not correlated ($\tau = -1$ indicates a perfect anti-correlation
while $\tau = 1$ indicates a perfect positive correlation).  For the solid points in
Fig.~\ref{fig:cor_shape}, Kendall's $\tau = -0.905$ with a probability $P < 0.004$ that the data are uncorrelated.  For the open points in Fig.~\ref{fig:cor_shape}, 
we find that $\tau = -0.810$ with a probability $P < 0.011$ that the data are uncorrelated.  
We interpret these trends as confirmation that the strengths and shapes of the silicate
features are correlated. Though crystalline silicate emission bands will also influence the
shape of the 10~$\mu$m silicate band, the correlations of
Fig.~\ref{fig:cor_shape} can be best explained by a change in the typical size
of the emitting dust grains. The dashed line shows the expected correlation for
an amorphous silicate of olivine stoichiometry with a grain size changing
continuously from 0.1~$\mu$m (lower right) to 2~$\mu$m (upper left). The only
exception is the 8.6 over 9.8~$\mu$m ratio of RX~J1612.6-1859A (\#3), where the
ratio is dominated by an unusually strong  emission from enstatite and silica. 

\begin{figure*}[t]
\begin{center}
\resizebox{0.9\hsize}{!}{\includegraphics[angle=0]{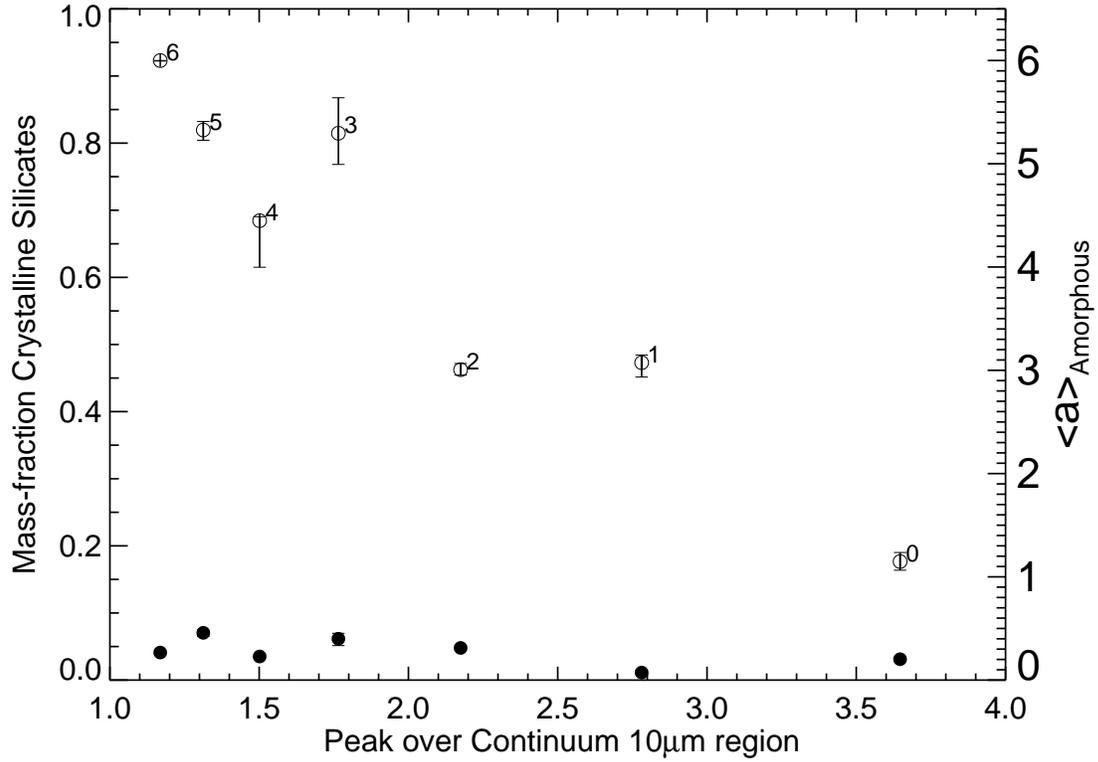}}
  \caption{Correlation between the derived mass fraction of crystalline silicates 
(filled symbols; left axis) and the peak-over-continuum ratio of the 10~$\mu$m silicate 
band and the mass averaged grain size of the amorphous silicates (open symbols; right axis) emitting at this wavelength region. As one can see, as the amorphous grains become bigger the 10~$\mu$m silicate band
 becomes weaker. No correlation between the mass fraction of crystalline silicates and the 
typical grain size of the amorphous silicates can be observed.
}
   \label{fig:cor_size_mass}
\end{center}
\end{figure*}

Fig.~\ref{fig:cor_size_mass} shows the relation between the mass weighed grain
size of the amorphous iron-magnesium silicates (open symbols; right axis) as
derived from our spectral decomposition of 0.1, 1.5 and 6~$\mu$m grains and the
peak over continuum value of the 10~$\mu$m silicate band. 
Examining this relation in more detail, we find a Kendall's $\tau$ of -0.81 for the rank ordering of peak-to-continuum emission compared to the mean size of amorphous silicate grains based on the analysis presented in Table~\ref{tab:t2_fs:fits}. The probability of
these variables not being correlated is 0.011. This correlation demonstrates that the peak to continuum ratio is correlated to the estimated size of the emitting amorphous iron-magnesium silicate grains. As a note of caution, recall that the exact values for the grain sizes derived are linked to the grain model used in the spectral decomposition. For reasons discussed in Section~\ref{sec:model}, we use homogeneous spheres to model the
amorphous silicates.  More complex grain structures, like fractal high-porosity
grains give qualitatively similar results \citep[e.g.][]{voshchi2006, min2006},
but require much larger aggregates to reproduce the observed band strengths.

\subsection{Coagulation and dust settling}
\label{sec:settling}

Having explored grain growth for our sample of T Tauri disks compared to the ISM
grain population, we now ask what effect coagulation could have on the disk
structure itself. As we do not have spatially resolved observations, we use the
shape of the SED to infer the geometry of the disks. It is long known that the
disk geometry strongly influences the shape of the SED
\citep[e.g.][]{kenyon1987}. Strongly flaring disks, intercepting a substantial
fraction of the radiation from the central star at large radii, show a rising
SED peaking at around 100~$\mu$m. 'Flattened' disks, on the other hand,
intercept far less of the radiation from the central star, and show a
power-law-like SED decreasing with wavelength. \cite{HerbigOverview} showed that
the disks in HAEBE systems can be divided into two groups based on the shape of
their SEDs. \cite{HerbigOverview} interpret this bi-modal behaviour of the SED in
terms of a bi-modal disk geometry, having either a flaring or a 'flattened' disk
structure. Recent spatially resolved observations seem to confirm the
interpretation that the different SEDs can be linked to different disk
geometries \citep[e.g.][]{leinert2004}. As an explanation for the different disk
geometries two different models have been proposed: (1) grain growth and
gravitational settling towards the disk mid-plane, and (2) 'self shadowing' of
the disk surface at larger radii by the inner ($\sim0.1$~AU) disk.

Coagulation models show that if the dust grains in the upper layers of a flared
disk become sufficiently large, they will gravitationally settle towards the
mid-plane of the disk, resulting in a flattened disk geometry
\citep[e.g.][]{Schraepler2004,nomura2006}. 
Recent model calculations concerning
the effect of grain coagulation and settling on the SED show that the changing
dust geometry from a flaring towards a flattened geometry due to the grain
settling results in SEDs consistent with observations
\citep[][]{allessio2006,dullemond_settling2004,Furlan2005,Furlan2006}

As an alternative, \cite{dullemond2004} explain the variation in observed SED
shapes for HAEBE and TTS systems as differences in disk heating caused by
differences in the self-shadowing of the disk surface. If the central star can
not directly irradiate the outer disk surface due to a puffed-up inner disk
(causing the surface of the outer disk to lay in the shadow of the inner disk),
the temperature and thus the scale height of the shadowed region will be
substantially lower compared to a disk with a directly irradiated surface. This
lower temperature will result in a smaller pressure scale height of the disk and
thus in a 'flattened' geometry.

\begin{figure*}[t]
\begin{center}
\resizebox{0.5\hsize}{!}{\includegraphics[angle=0]{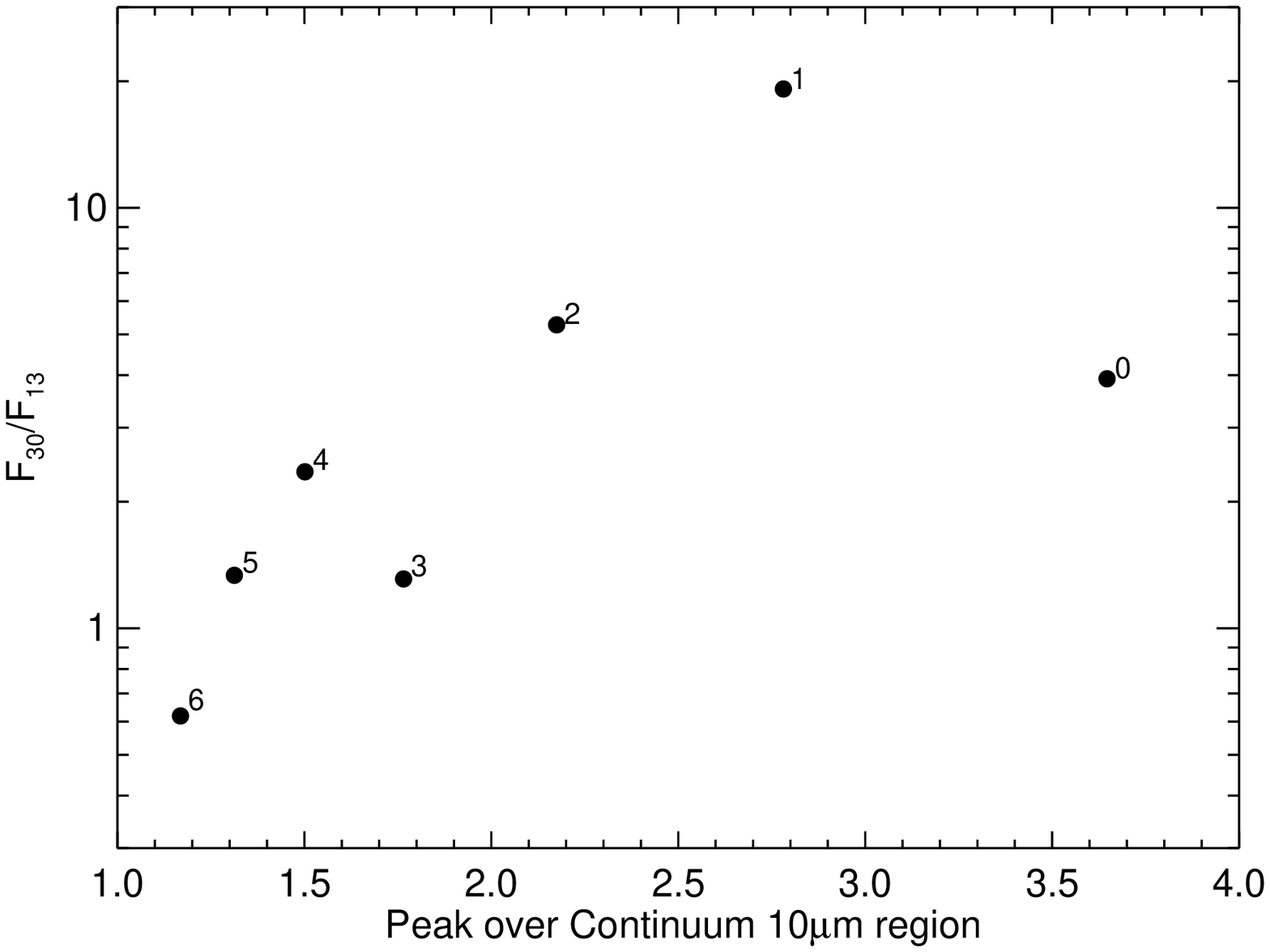}}
\resizebox{0.5\hsize}{!}{\includegraphics[angle=0]{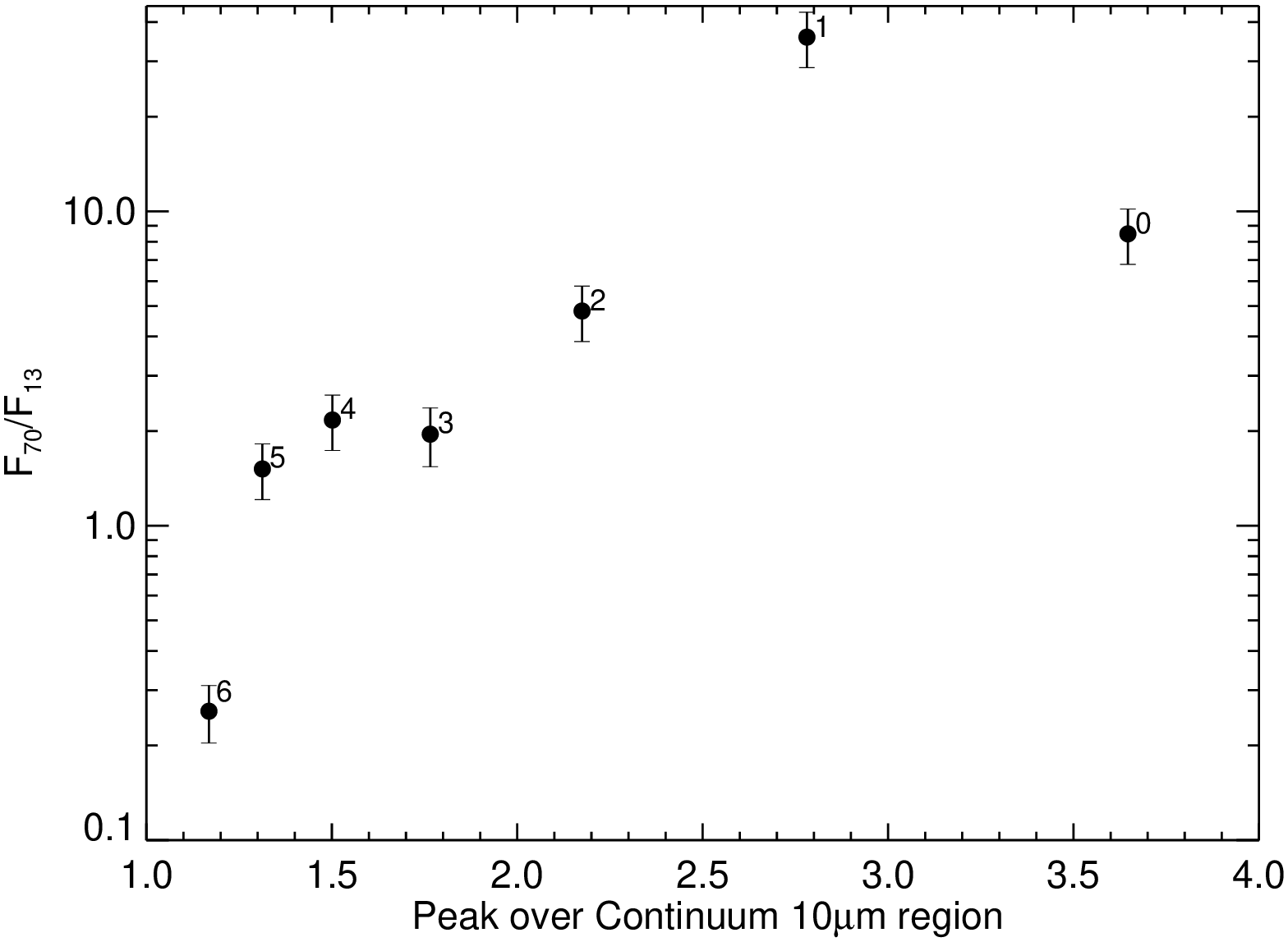}}
\caption{Correlation between the far-IR slope of the spectral energy distribution, a measure for disk geometry, and the peak-over-continuum ratio of the 10~$\mu$m silicate band, a measure for the typical grain size. The far-IR slope is measured  by the ratio of the flux at 30~$\mu$m over 13~$\mu$m (upper panel) and the 70~$\mu$m over 13~$\mu$m flux ratio (lower panel). A clear correlation can be seen: As the 10~$\mu$m silicate band becomes weaker, the slope of the 
spectral energy distribution decreases. The 13 and 33~$\mu$m fluxes are synthetic photometry points derived from the \emph{Spitzer} low-resolution spectra, the 70~$\mu$m fluxes are photometric data from the MIPS instrument on-board  the \emph{Spitzer Space Telescope}. Note that the error bars in the upper
panel are the same size as the symbols and reflect the small internall uncertainties in our IRS spectra.
The larger error bars in the lower panel mainly reflect the uncertainties in the absolute flux calibration between the MIPS and IRS instrument.
}
   \label{fig:cor_slope}
\end{center}
\end{figure*}

A way to test these models is to look for a relation between the measured grain
sizes and the geometry of the disk. In Fig.~\ref{fig:cor_slope} we explore the
relation between the amorphous silicate grain sizes, as measured by the strength
of the 10~$\mu$m silicate feature, and the disk flaring angle, as measured by
the 30~$\mu$m over 13~$\mu$m (upper panel) and the 70~$\mu$m over 13~$\mu$m 
(lower panel) flux ratios. Here, the 13~$\mu$m and 30~$\mu$m fluxes are
synthetic photometry points derived from our IRS spectra using an
1~$\mu$m, respectively, 2~$\mu$m wide box centered on the quoted wavelengths. 
The 70~$\mu$m fluxes are derived from MIPS observations. 
 Fig.~\ref{fig:cor_slope} provides quantitative evidence that these variables are
indeed correlated. The Kendall's $\tau$ for the top panel 
is 0.7 with $P = 0.03$, while for the bottom panel $\tau = 0.81$ with
$P = 0.01$, both indicating a significant positive correlation. 
As we saw in Fig.~\ref{fig:cor_size_mass}, the peak-to-continuum ratio of the 10~$\mu$ silicate features is strongly anti-correlated with grain size.  As a result, we
conclude from  Fig.~\ref{fig:cor_slope} that grain size is strongly correlated with
SED slope.  Larger flaring angles of the disk produce
larger flux ratios because of enhanced emission from cooler grains further out
in the disk that are exposed to the stellar flux by the flaring. With increasing
grain size (decreasing 10~$\mu$m silicate band) the flaring angle of the disk
appears to decrease (as traced by the slope of the SED towards longer
wavelengths). Our results are in qualitative agreement with the model
predictions of \cite{allessio2006} and \cite{dullemond_settling2004}, and
provide direct spectroscopic evidence for the link between the typical size of
the dust grains and the disk structure. These results suggest that in TTS
systems it is coagulation and grain settling towards the mid-plane which
determines the disk geometry and thus the SED, rather then self-shadowing by the
disk.

\subsection{The crystalline silicates in protoplanetary disks}
\label{sec:crystalline}

\begin{figure*}[t]
\begin{center}
\resizebox{0.6\hsize}{!}{\includegraphics[angle=0]{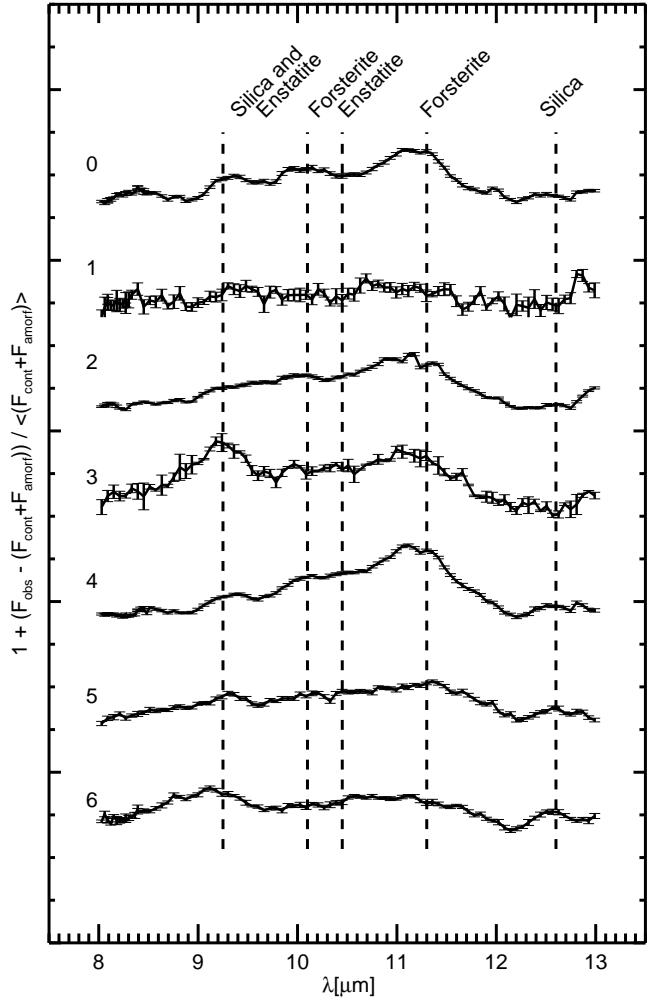}}
  \caption{The emission bands of crystalline silicates in the 10~$\mu$m spectral
window. Shown are the Spitzer low-resolution spectra normalized to the fitted
amorphous silicate, PAH, and continuum model (see also Section~\ref{sec:model}).
The normalized spectra all have the same vertical scale between 0.97 and 1.15
but are offset for clarity. The spectra are ordered from top to bottom by
decreasing 10~$\mu$m silicate band strength. The ID numbers correspond to those
listed in Table~\ref{tbl:stars}. Also indicated in this figure are the positions
of the main emission bands of silica, forsterite, enstatite dust grains. }
  \label{fig:young_disks_short_no_amorf}
\end{center}
\end{figure*}

\begin{figure*}[t]
\begin{center}
\resizebox{0.6\hsize}{!}{\includegraphics[angle=0]{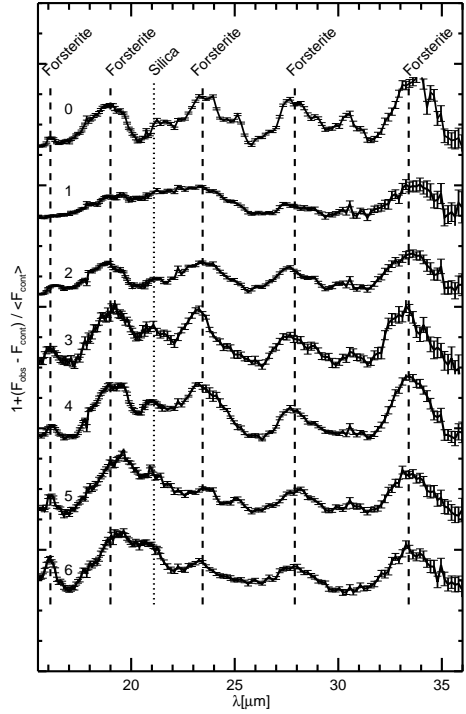}}
  \caption{Continuum normalized spectra of the FEPS TTS sample. Shown are the
Spitzer low-resolution spectra between 17 and 36~$\mu$m normalized to the fitted
continuum using a low order polynomial (see also Section~\ref{sec:model}). The
normalized spectra all have the same vertical scale between 0.97 and 1.15 but
are offset for clarity. The spectra are ordered from top to bottom by decreasing
10~$\mu$m silicate band strength. The ID numbers correspond to those listed in
Table~\ref{tbl:stars}. Also indicated in this figure with the dashed and dotted
lines are the positions of the main spectral features of forsterite and silica.}
  \label{fig:young_disks_long_no_cont}
\end{center}
\end{figure*}

The analysis of features produced by crystalline silicates  provides important
information on dust processing in protoplanetary disks. Presently, no
crystalline silicates have been observed in the diffuse ISM. \cite{kemper2005}
place an upper limit of 1\% on the crystalline silicate mass fraction of the
diffuse ISM. The much higher mass fractions observed in protoplanetary disks 
\citep[e.g.][$\sim$5\% in this study]{processing,Roy2005}
therefore, imply that the observed crystalline silicates had to be formed in the
disks themselves, rather then being incorporated directly from the ISM into the
disk \citep[see][]{henning2005}. This makes crystalline silicates a 
tracer of dust processing and evolution in protoplanetary disks.
Figs.~\ref{fig:young_disks_short_no_amorf} and
\ref{fig:young_disks_long_no_cont} show the observed crystalline silicate bands
at mid- and far-IR wavelengths, respectively, for our TTS sample. As described
in Section~\ref{sec:analysis} the spectra have been normalized and, for the 10 $\mu$m
regime, corrected for the amorphous silicate contribution. These normalized
spectra clearly show the emission bands characteristic of emission from the
crystalline dust species forsterite, enstatite and silica. Using the opacities
plotted in Figs.~\ref{fig:kappa_short}a to \ref{fig:kappa_short}c, these three
crystalline dust species, in combination with the amorphous silicates, produce an
excellent fit to the Spitzer spectra as shown in Fig.~\ref{fig:fit}. 
The overall mass fraction of crystalline silicates based on the spectral decomposition
of the 10 $\mu$m silicate band is around $\sim$5\% (see also Table~\ref{tab:t2_fs:fits}).
We find no conclusive evidence for other crystalline silicates.
Interestingly, other crystalline dust species, like iron
sulfides or oxides can be expected to exist in protoplanetary disks
\citep[e.g.][]{pollack1996}. Looking at Figs.~\ref{fig:fit} and
\ref{fig:young_disks_long_no_cont}, weak features at 25.5~$\mu$m (source
number 0 and 5) and 30.5~$\mu$m (source number 0 and 3) can be observed which are
not properly reproduced by our dust model. These features, however,
are 1) not conclusively seen in the other targets; 2) the bands do not coincide with
known bands of materials like iron-sulfide or iron-oxide \citep{begemann1994,henning1995};
and 3) no clear relation between these weak bands can be observed. 
The possibility that these very weak bands are an
instrumental artifact in the form of some residual fringing can not be ruled out
at this stage.

\subsubsection{Crystallinity and grain size}

\begin{figure*}[t]
\begin{center}
\resizebox{0.7\hsize}{!}{\includegraphics[angle=0]{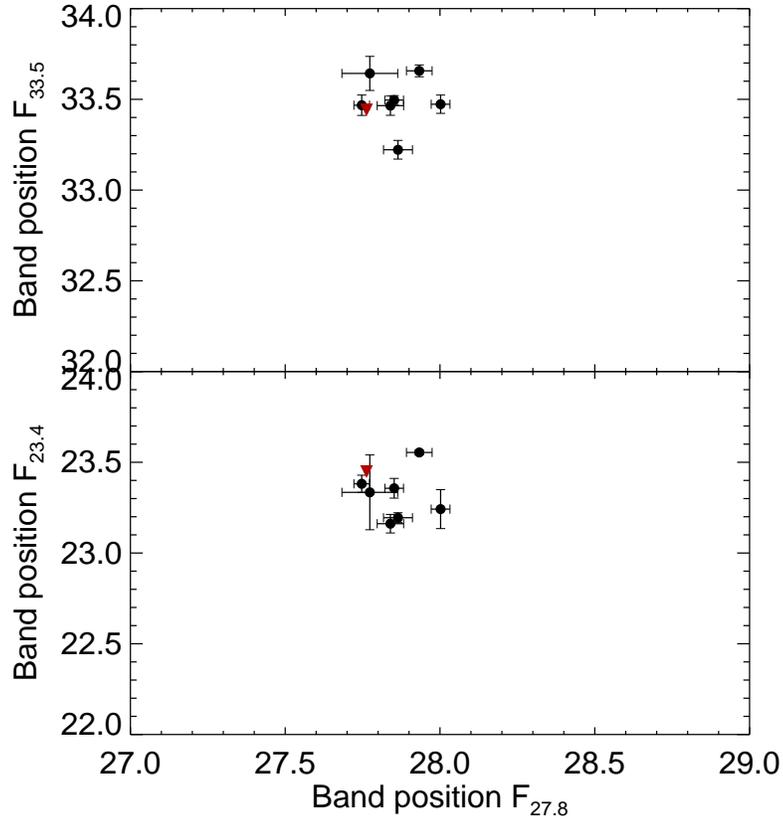}}
\caption{Comparison between the forsterite band position at 27.8~$\mu$m and that
of the 33.5~$\mu$m feature (top panel) and the 23.4~$\mu$m feature (lower
panel). Also indicated are the calculated nominal band positions for 0.1~$\mu$m
forsterite grains (filled triangles). }
  \label{fig:cor_band_pos}
\end{center}
\end{figure*}

\begin{figure*}[t]
\begin{center}
\resizebox{0.8\hsize}{!}{\includegraphics[angle=0]{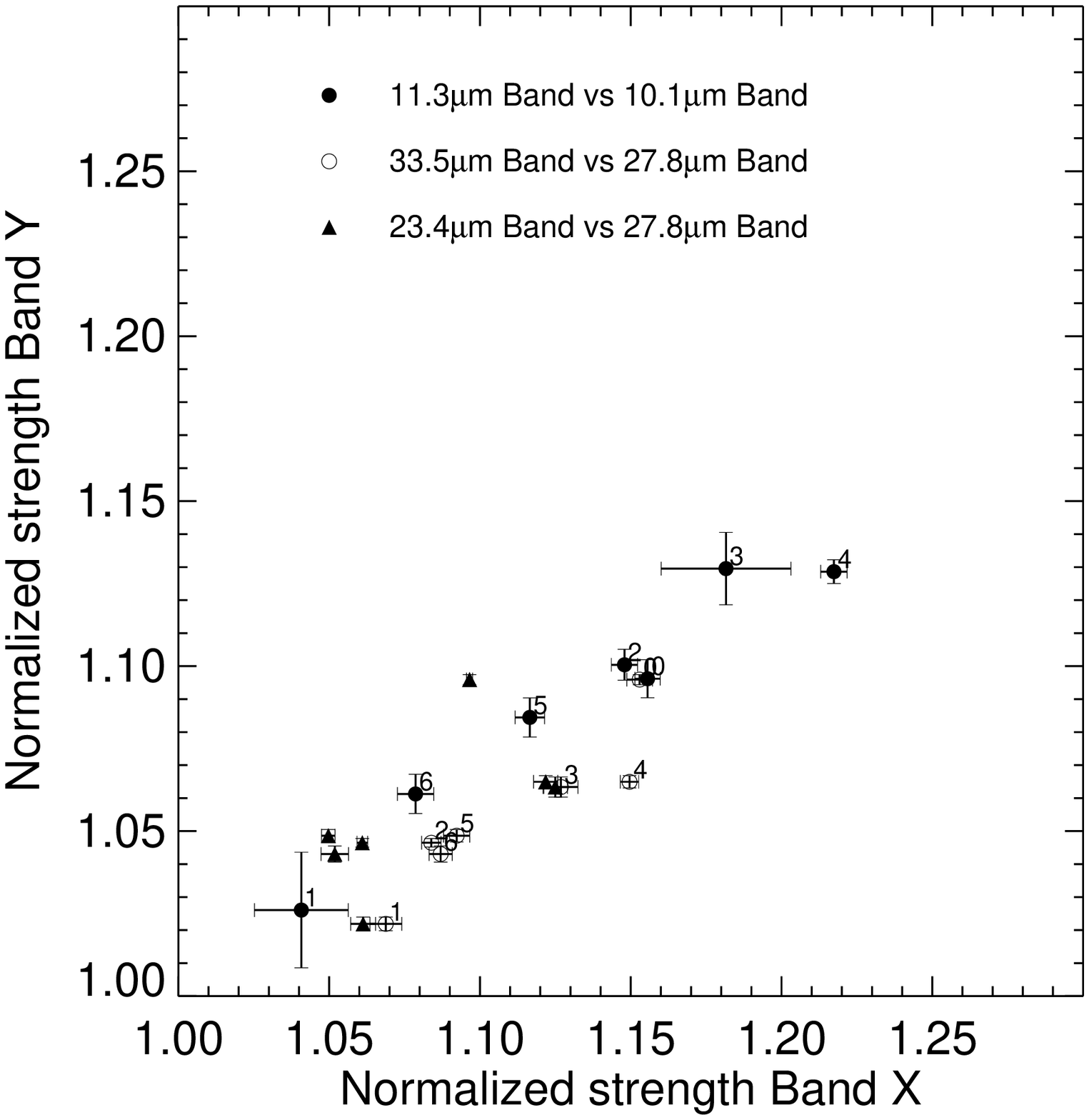}}
  \caption{Comparison between the measured strength of the forsterite bands at 11.3~$\mu$m (plotted on x-axis) and 10.1~$\mu$m (filled circles), between the 33.5~$\mu$m (plotted on x-axis) and the 27.8~$\mu$m spectral features (open circles) and the 23.4~$\mu$m (plotted on x-axis) and the 27.8~$\mu$m spectral features (filled triangles).}
  \label{fig:cor_bands}
\end{center}
\end{figure*}

As for amorphous dust grains, the shape, strength and wavelength positions of
the emission bands of the crystalline dust species yield information on the
typical size of the emitting dust grains. As the crystalline silicates have many
narrow spectral features sensitive to shape, size and composition of the dust
grains, an excellent determination of the grain properties can be made. Using
the analysis method described in Section~\ref{sec:analysis},
Figs.~\ref{fig:cor_band_pos} and \ref{fig:cor_bands} show the correlations
between the measured position and strengths of the main forsterite resonances in
our TTs sample.  No trend in the band positions can be observed, which hardly vary from the nominal position expected for 0.1~\micron~ sized grains (filled triangles). This is in sharp contrast to the results for the amorphous silicate component, where grain sizes of at least 6~\micron~ are required to reproduce our observations. Also,
the band ratios between the main forsterite resonances do not substantially vary
from source to source. By comparing the strengths of neighbouring bands, which minimizes the influence of possible source to source variations in the mass over temperature distribution, a linear relation can be observed. This relation reflects differences in the crystalline mass fractions in the different sources, rather than differences in the grain size of the crystalline silicates. A Kendall's $\tau$ test for the data plotted in Fig.~\ref{fig:cor_bands} results in $\tau =0.81$ with $P=0.011$ for the  11.3~$\mu$m versus 10.1~$\mu$m band strength, $\tau =0.90$ with $P=0.004$ for the  33.5~$\mu$m versus 27.8~$\mu$m band strength, and $\tau =0.24$ with $P=0.45$ for the  23.4~$\mu$m versus 27.8~$\mu$m band strength. The apparent lack of correlation between the 23.4~$\mu$m and 
27.8~$\mu$m band strength most likely reflects the difficulty in fitting an accurate local continuum underneath the 23.4~$\mu$m band. Still, those sources showing the strongest 23.4~$\mu$m bands also show the strongest bands at 27.8~$\mu$m.

\begin{figure*}[t]
\begin{center}
\resizebox{0.9\hsize}{!}{\includegraphics[angle=0]{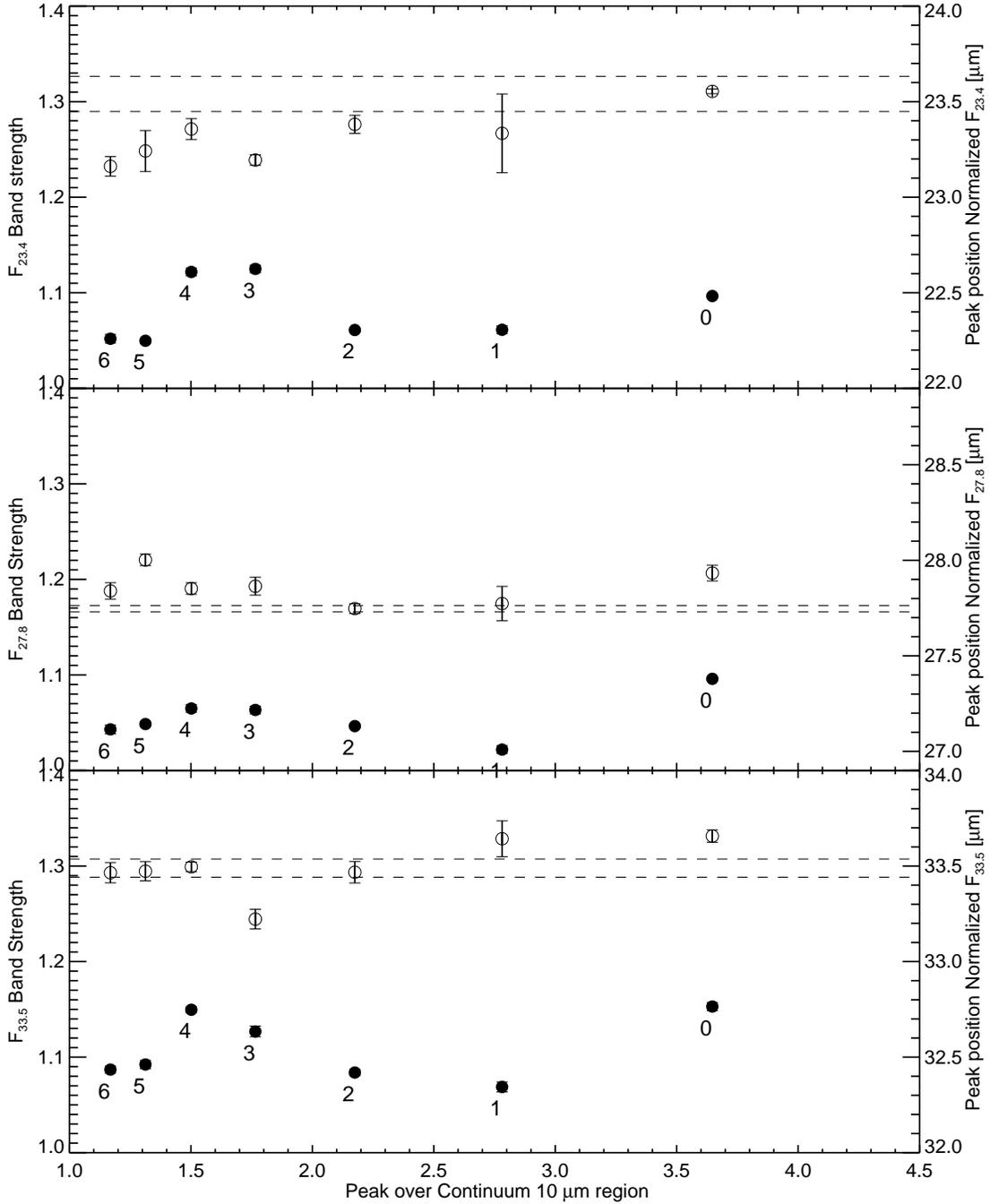}}
  \caption{Correlations between the peak over continuum ratio of the 10~$\mu$m
silicate band and the band strengths (filled symbols; left axis) and positions
(open symbols; right axis) of three spectral emission bands of crystalline
silicates. The top, middle and bottom panels show the correlations for the
23.4~$\mu$m, 27.8~$\mu$m, and the 33.5~$\mu$m band, respectively. As one can
see, both position and normalized band strength of these crystalline silicate
bands show almost no variations and are not correlated to the silicate emission
at 10~$\mu$m. }
   \label{fig:cor_xbands}
\end{center}
\end{figure*}

From the spectral decomposition in the 10~$\mu$m wavelength region (see also
Table~\ref{tab:t2_fs:fits}), we determined the relative mass fraction of
crystalline silicates contributing to the crystalline emission bands.
Fig.~\ref{fig:cor_size_mass} also shows the relation between
the observed amorphous silicate grain size and the crystalline silicate mass
fraction. A Kendall's $\tau$ test gives $\tau = -0.43$ with $P = 0.18$ suggesting reasonable probability that the crystallinity and grain size of the (bulk) amorphous dust are uncorrelated. Determining the mass
fraction of crystalline silicates depends on how well one can determine the
mass fraction of the amorphous dust component. Within the 10~$\mu$m spectral
region, probing the warmest dust, this poses no problem as the amorphous
silicates show a clear spectral signature. At the longer ($\sim$20~$\mu$m)
wavelengths, probing the colder dust at larger disk radii, however, amorphous
silicates lack clear spectral signatures, which makes it difficult
to determine the physical properties of the
colder grains. However, the strengths of the emission bands over the local
continuum in the $\sim$20 - 30~$\mu$m spectral region, still provide a measure
of the relative crystalline mass fraction as a function of wavelength, and thus
as a function of temperature and hence as a function of radial distance in the disk. Fig.~\ref{fig:cor_xbands} compares the strength of the 10~\micron~ silicate band -a measure of the typical grain size of the amorphous silicates- to the observed band strengths and positions of three of the main crystalline silicate bands. 
No conclusive correlation between the amorphous and
crystalline bands can be observed, implying that both the observed crystalline mass fraction and size of the crystals are not correlated with the grain size of the amorphous
dust, and thus grain growth and crystallization appear to be unrelated.
A Kendall's $\tau$ test for the relation between the 
peak-over-continuum value versus the strength and position of the 27.8~$\mu$m band results in $\tau =0.147$ with $P=0.65$ and $\tau =-0.048$ with $P= 0.88$ , respectively.
For the  the relation between the peak-over-continuum value versus the strength and position of the 33.5~$\mu$m band we find $\tau = 0.048$ with $P=0.89$ and $\tau = 0.52$ with $P= 0.1$, and for the relation between the peak-over-continuum value versus the strength and position of the 23.4~$\mu$m band we find $\tau =0.33$ with $P=0.29$ and $\tau = 0.62$ with $P=0.051$, respectively. Similar to Fig.~\ref{fig:cor_bands}, discussed in the previous paragraph, the possible weak correlation we observe for the 23.4~$\mu$m band is most likely due to a systematic problem in correctly determining the local continuum using a polynomial fit. As we can see in Fig.~\ref{fig:specta}, the shape of the SED in the 20~$\mu$m region varies considerable. As we shown in Section~\ref{sec:settling}, there is a correlation between the shape of the SED and the peak-over-continuum value of the 10~$\mu$m silicate band. The possible correlation with the 23.4~$\mu$m band, therefore, might simply reflect a systematic difference in the local continuum estimate between source with a towards longer wavelengths rising SED and those for which the SED is decreasing.

At first glance the spectrum of a source like 1RXS~J161410.6-230542
(lower left panel in Fig.~\ref{fig:fit}) seems to exhibit a much higher
crystalline mass fraction than a source like RX~J1842.9-3532 (upper left panel
Fig.~\ref{fig:fit}). The apparently more pronounced crystalline silicate bands
in the former TTS system, however, do not reflect a larger crystalline mass
fraction but rather less amorphous silicate emission.  As the amorphous grains
grow with respect to the crystalline silicates, their relative opacities
decrease and hence their contribution to the observed spectra diminishes. It is,
therefore, crucial to have an accurate determination of the grain size of the
(bulk) amorphous silicates. If we would have used a smaller maximum grain size
in our spectral decomposition, we would have estimated that no amorphous
silicates are present in the 1RXS~J161410.6-230542 system, leading to a 100~\%
crystalline silicate mass fraction. This would, consequently, have led to a
relation between the grain size of the amorphous dust and the crystallinity, as
was noted by \cite{Roy2005}. This result could be
an artifact from underestimating the maximum grain size in their ground based observations, leading to a lower mass estimate of the amorphous dust component. 
Compared with these ground-based studies, 
we should note that the \emph{Spitzer} data have the advantage to allow for a
determination of the continuum outside of the spectral window accessible from
the ground. This results in a better characterization of the strength of the
10~$\mu$m silicate band, and thus of the grain size, especially in the case of
very weak emission features.

\subsubsection{The crystallization process}

\begin{figure*}[t]
\begin{center}
\resizebox{0.5\hsize}{!}{\includegraphics[angle=0]{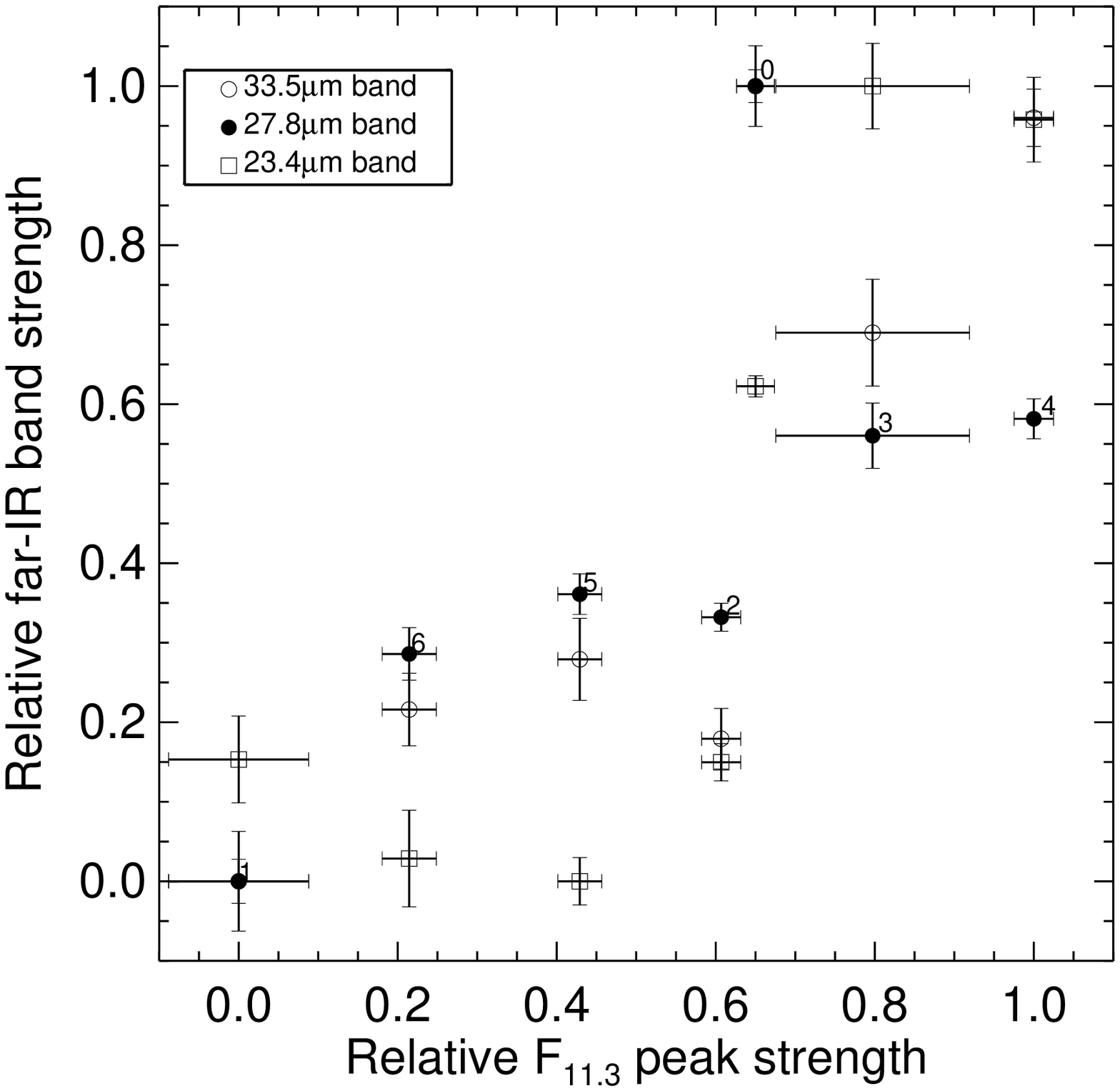}}
\resizebox{0.5\hsize}{!}{\includegraphics[angle=0]{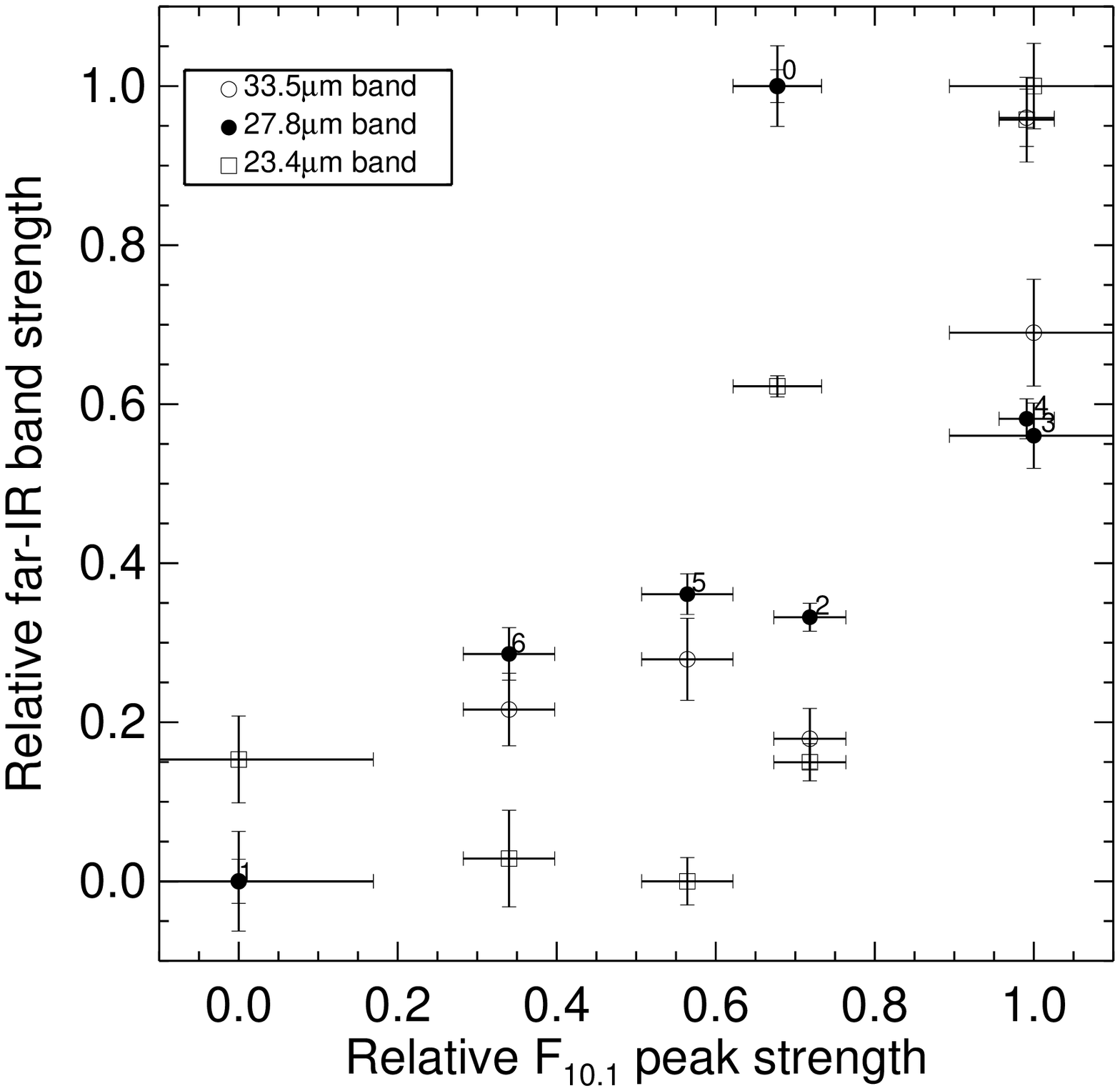}}
  \caption{Correlations between the measured band strengths of the crystalline
silicate features in the 20 to 35~$\mu$m wavelength region and those observed in
the  10~$\mu$m spectral region. The top figure shows the comparison between the
band strength of the 11.3~$\mu$m and that of the 33.5~$\mu$m spectral feature
(open circles), the 27.8~$\mu$m band (filled circles) and 23.4~$\mu$m band (open
squares), respectively. The lower figure shows a similar comparison but with the
band strength of the 10.1~$\mu$m spectral feature.}
  \label{fig:cor_far_ir_bands}
\end{center}
\end{figure*}

Apart from the comparison between the crystalline  and the amorphous emission
bands, the inter-comparison between the crystalline spectral features can
provide crucial information. Though we are not spatially resolving the
protoplanetary disks, from model calculations of the temperature distribution in
the disk around a typical TTS, one can make an estimate of the size of the
emitting region at a given wavelength \citep[e.g]{Kessler_size}. 
\cite{Kessler_size} argue that the silicate feature at around 10~$\mu$m
originates from the disk surface at round 1~AU from the central star. The
spectral features observed at the longest wavelengths, at around 30~$\mu$m,
correspond to a temperature of $\sim$120~K; the temperature at which a black
body would emit most of his radiation at the given wavelength. Assuming a
$\lambda^{-1}$ dependency for the dust opacity, and a typical stellar
temperature and radius of 5500 K and 1 $R_{\odot}$, respectively, dust grains
attain a temperature of 120 K at a radius of about 15~AU from the central star.
The implies that with the IRS instrument we can probe the inner $\sim$15~AU of a
disk in a TTS system. The dust composition of the disks outward of $\sim$15~AU
can not be determined as the dust grains at these large radii are to cold to
emit at wavelengths covered by the IRS spectrograph. Comparing the relative band
strengths of forsterite in the 10~$\mu$m wavelength region probing the warmest
dust, with the forsterite bands at longer wavelengths probing the colder dust
component further out in the protoplanetary disks, a possible trend can be observed (see
Fig.~\ref{fig:cor_far_ir_bands}). A Kendall's $\tau$ test for the relation between the 11.3~$\mu$m peak strength and the peak strengths of the 23.4, 27.8 and 33.5~$\mu$m bands results in  $\tau = 0.52$  with  $P=0.1$,  $\tau = 0.71$ with $P=0.024$ and $\tau = 0.62$ with $P = 0.051$, respectively.

On average, those systems showing the strongest forsterite
emission bands in the 10~$\mu$m spectral region also show the strongest bands at
the longer wavelengths. One could interpret this possible correlation 
as evidence of a global crystallization process, where the crystallinity of the inner $\sim$15~AU of the disk is increased from non-detectable ISM values to the observed mass fractions of about $\sim$5\% as measured in the 10~$\mu$m spectral region. Whether this
means that the crystalline silicates are formed throughout the entire inner
disk, or formed locally and then redistributed throughout the inner disk will be
discussed in Section~\ref{sec:discussion}.   One should note that due to the
uncertainties in determining the amorphous mass fractions in a quantatitive way
at the longer wavelengths (see also Section~\ref{sec:model}), the radial density
profile of the crystalline silicates remains uncertain. It is possible that in
colder parts of the disk, at larger radial distances, the fraction of
crystalline grains deviates from that determined for the inner parts. However,
our results show that, irrespective of the exact radial profile, the
crystallinity of the very inner ($\sim$1~AU as measured at 10~$\mu$m) and outer
($\sim$5-15~AU as measured at wavelengths $\sim20-30$~$\mu$m) disk regions are
related.

\begin{figure*}[t]
\begin{center}
\resizebox{\hsize}{!}{\includegraphics[angle=0]{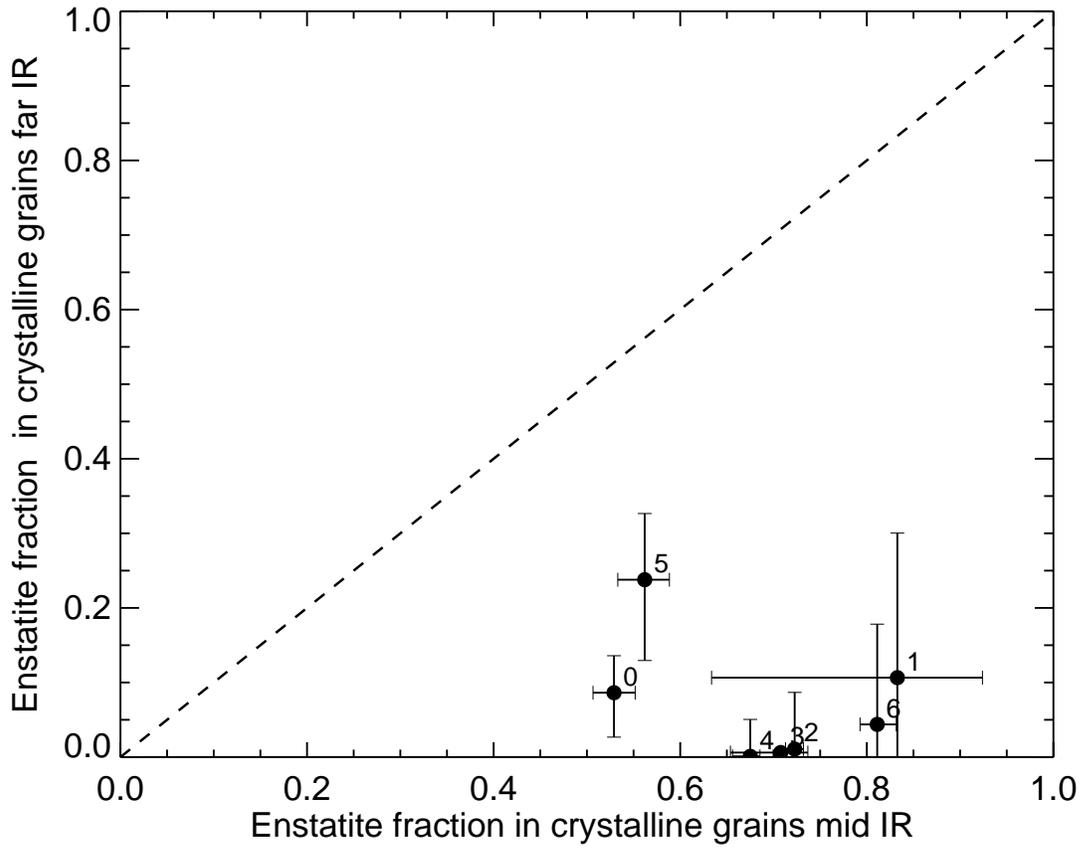}}
  \caption{Correlation between the enstatite mass fraction of the crystalline silicates as
   measured in the 10~$\mu$m spectral window and the enstatite mass fraction of the crystalline silicates
   derived from analyzing the longer wavelength features between 20 to 35~$\mu$m.}
  \label{fig:cor_xsil_mid_far_ir}
\end{center}
\end{figure*}

A similar comparison for the enstatite emission bands is difficult as at the
longer wavelengths our spectra show no features which can be uniquely identified
with enstatite. Comparing the strength of the main enstatite band at 9.2~$\mu$m
and the longer wavelength bands of forsterite we find no clear correlation.
 This suggests that both crystalline species are to a certain extent
unrelated, and a general increase of forsterite abundance seems not to correspond to a
similar increase in enstatite abundance. Although the crystalline silicate mass relative to
that of the amorphous silicates can not be determined at the longer wavelengths, 
our spectral decomposition as a function of wavelength, and thus radius of the emitting regions,
can determine the mass ratios between the different crystalline silicates. In Fig.~\ref{fig:cor_xsil_mid_far_ir} the relation
between the enstatite fraction of the crystalline grain populations as measured
within the 10~$\mu$m wavelength region, 
and that determined from the longer wavelength spectral regions, is
plotted. Surprisingly, these results imply that the mass fraction of enstatite
varies in the radial direction, with a larger enstatite fraction in the warmer
inner disk than in the colder outer ($\sim$5 to 15~AU) regions. Another
difference between the forsterite and enstatite grain population is the typical grain size.
The grain size  of the enstatite grains ($\sim$1~$\mu$m)
is larger than that of the forsterite grains ($\sim$0.1~$\mu$m). Again this
points towards differences in the formation mechanism and/or conditions
producing these crystalline species. The implications of the above results will
be discussed in Section~\ref{sec:discussion}.

\subsection{PAH emission in protoplanetary disks around T~Tauri stars}
\label{sec:pah}
\begin{figure*}
\begin{center}
\resizebox{0.8\hsize}{!}{\includegraphics[angle=0]{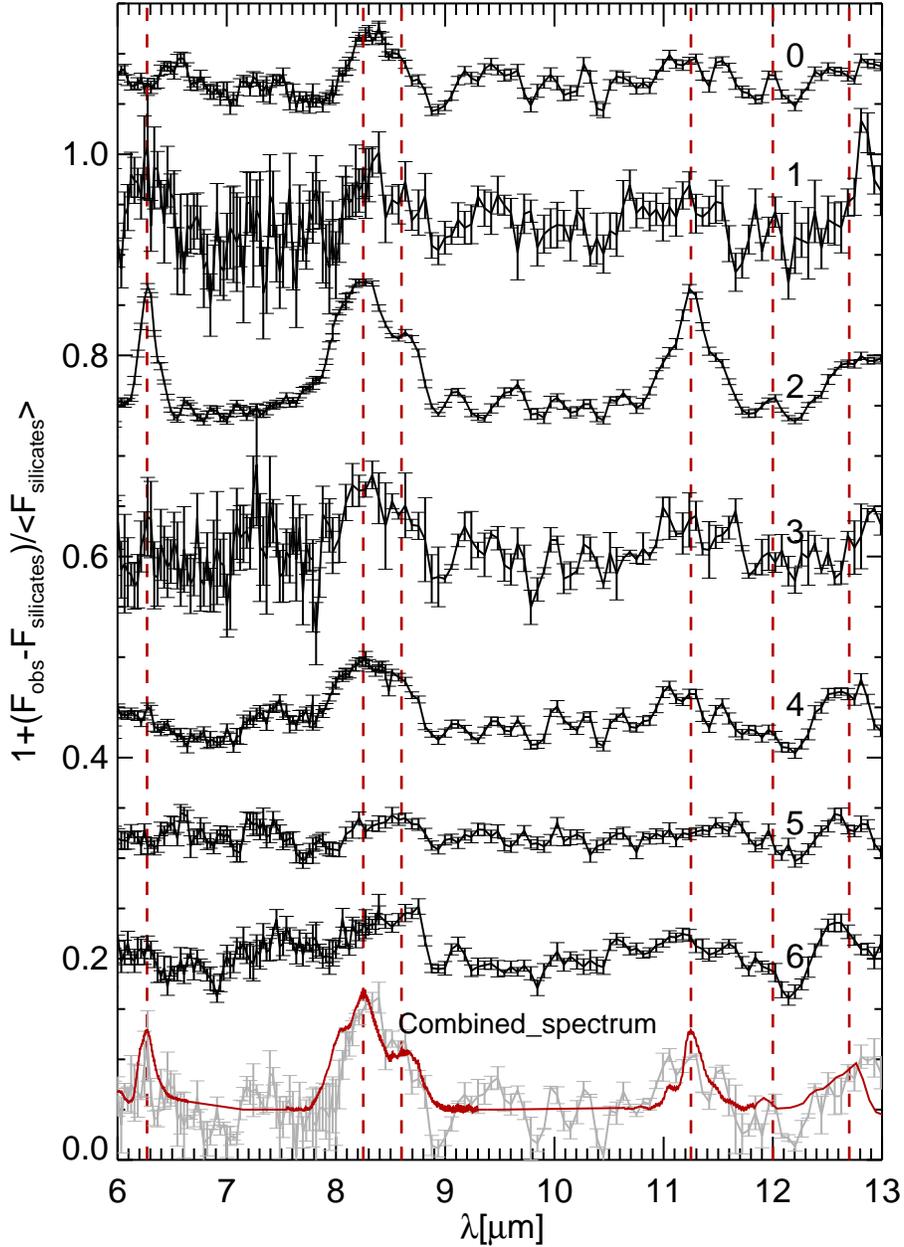}}
  \caption{Observed PAH emission bands in the spectra of the FEPS TTS sample.
Shown are the Spitzer low-resolution spectra normalized to the fitted silicate
model (see also Section~2.2 and Table~\ref{tab:t2_fs:fits}). The first 5 spectra
clearly show an emission feature at 8.2~$\mu$m we assign to emission from PAH
molecules. Shown at the bottom of the figure is the average spectrum of source
\# 0,1,3 and 4, over-plotted with the PAH template spectrum used in our fitting
procedure. The dashed lines mark the positions of the main PAH bands. }
   \label{fig:pah}
\end{center}
\end{figure*}

So far, little is known about the presence of PAH molecules in disks around
low-mass pre-main sequence stars. Systems with a relatively strong UV field can
readily excite PAH molecules, as is observed for the more massive and luminous
Herbig~Ae/Be systems where clear PAH emission signatures can be seen \citep[e.g.][]{HerbigOverview,sloan2005}. 
Though observations of UV poor reflection nebula and PAH 
laboratory measurements and models \citep[][]{Li2002,Mattioda2005}
show that longer wavelength photons can also excite PAH molecules, little
evidence has been found for IR emission bands from stochastically heated
molecules or very small grains from the circumstellar disks around low-mass
stars as studied in this paper. After carefully fitting our silicate model, a
residual feature centered at 8.2~$\mu$m could be seen in the model subtracted spectrum of
five of the observed targets, which we tentatively identify
with emission from stochastically heated PAH molecules. Fig.~\ref{fig:pah} shows
the normalized residual spectrum after subtraction of the best-fit silicate model. 
The strongest bands can be seen in the residual spectrum of HD~143006 (\#2), the
most luminous source in our sample (G6/8 spectral type), 
which also shows clear evidence for emission bands at 6.2~\micron~ and 11.2~\micron. 

As the PAH molecules require stellar photons to be excited, the molecules
contributing to the emission bands have to be located at the optically thin
surface layers of the disk, in direct view of the central star. As many papers
have shown for the Herbig~Ae/Be systems \citep[e.g.][]{HerbigOverview, acke_iso_spec2004}, 
there exists a direct correlation between the presence of PAH emission bands and the shape of
the SED, a measure of the flaring angle of the disk. Those disks having an SED
consistent with a flattened disk structure, and hence a less illuminated disk
surface show little or now evidence of PAH emission. These results are also
confirmed by spatially resolved spectroscopy showing the PAH emission bands to
come from the flaring surface at larger radii of the disk
\citep[][]{vanboekel2004}. We can observe a similar trend, where those sources
having SEDs consistent with more flattened disk structures show no clear
evidence for the 8.2~$\mu$m feature. This apparent correlation between the lack
of the 8.2~$\mu$m feature and the flaring angle of the disk, therefore, suggests
that this emission band is not due to an unidentified dust species in thermal
equilibrium, but rather a  stochastically heated large molecule or very small
grain species at the disk surface.

Surprisingly, the observed band in the 8~$\mu$m region is not centered between
7.7 to 7.9~$\mu$m, where PAH emission is usually observed. The band position at
8.2~$\mu$m has only been  measured in the spectra of a few other sources.
\cite{peeters2002} observed this band position in two post-AGB stars, and
\cite{sloan2005} observed the feature in the HAEBE system HD~135344. The
observed bands in these three objects, however, show a slightly broader profile
than is observed in the TTS systems. The observed relatively weak or absent
6.2~\micron~ and 11.2~\micron~ features with respect to the 8.2~\micron~ band is
consistent with the \cite{peeters2002} and \cite{sloan2005} studies. Further,
the HAEBE system from the latter study showing the 8.2~$\mu$m PAH band, has the
lowest UV-flux of the four systems discussed in that paper, which would be
consistent with our observations of lower luminosity systems. Presently, no
conclusive explanation for the carrier of the 8.2~\micron~ feature has been put
forward. We can only speculate that the chemistry, ionization state and/or
structure should be substantially different for our sources compared to the PAH
population in HAEBE systems. \cite{Li2002} argue that in an environment low in
UV where longer wavelength photons excite the PAHs, the PAH molecules have to be
ionized and/or large to be excited. The presumably much stronger X-ray emission
in the TTS systems with respect to their higher mass counterparts may play also
a role here, driving a different chemistry and ionization state. On the one
hand, X-ray photons may increase the electron abundance in the gas which would
reduce the PAH ionization. On the other hand, a strong X-ray field could
directly (multiple times) ionize the PAH molecules due to the Auger effect, and
thereby change them or even destroy the smallest. 

\section{Discussion}
\label{sec:discussion}

The circumstellar disks in T~Tauri star systems are believed to be the sites of
on-going planet formation, and thus represent an analogue for the proto-solar
nebula. As the disks evolve with time, the sub-micron sized dust grains present
at the formation time of the disks coagulate to form larger objects and
eventually planet(esimal)s \citep{Beckwith2000, henning2006}. 
By studying the characteristics and evolution of the
disk and its dust composition, valuable insights can be obtained into the
processes leading to the formation of planets, and important constraints on disk
and planet formation models can be derived. Also, by analogy, TTS systems 
can provide clues into the early evolution of the solar system.
In the following we will discuss the implications of our findings presented in the previous sections.
Note that our sample only spans a limited range in stellar parameters and consists of older, 
long surviving disks. Our conclusions are applicable to similar systems, but 
possible effects of stellar properties or age on the evolution
of circumstellar disks can not be address directly by this study. For this our results will
have to be compared to the disk properties of a larger sample, spanning a wide range in stellar properties and evolutionary stages.

\subsection{Grain growth and disk structure}
Based on our analysis of low-resolution infrared spectra of seven TTS systems
obtained with the Spitzer Space Telescope, we find clear evidence of processing
and growth of the silicate dust species present in their protoplanetary disks.
We interpret observed variations in the thermal emission from amorphous silicate dust
species as evidence for grain sizes which are substantially different from those
observed in the ISM, and argue for grain growth within the protoplanetary disks.
For the first time, we find a clear correlation between the strength of the
amorphous silicate feature, measuring the typical grain size of these
grains, and the shape of the mid-infrared spectral energy distribution,
measuring the disk scale height, i.e., flaring of the disk.  In the literature, two
possible explanations have been put forward to explain the differences in disk flaring:
First, self-shadowing of the disk due to an enhanced scale height of the very
inner parts. This prevents the central star from illuminating the surface of
the outer disk, required to sufficiently heat the disk for it to flare
\citep[][]{dullemond2004}. Second, grain growth and consequent gravitational
settling toward the disk mid-plane  \citep[e.g.][]{Schraepler2004,nomura2006,allessio2006,dullemond_settling2004,Furlan2005,Furlan2006}. 
Our results, linking an increase in grain size to a decrease of the disk flaring, 
clearly argue for the latter explanation. 

An earlier study by \cite{Apai2004}, studying the brown dwarf system CFHT-BD-Tau 4, also suggested 
a correlation between the strength of the silicate emission band and the shape of the SED.
\cite{acke_iso_spec2004} studied a sample of HAEBE systems observed with the
Infrared Space Observatory (ISO). Because no correlation could be found between
the observed silicate emission bands and the shape of the SED, the authors
concluded that their observations were consistent with the self-shadowing disk
model rather than coagulation and grain settling.  In apparent contradiction,
\cite{acke_mm_grainsize2004} showed a correlation between grain size, as
measured by the (sub) millimeter slope of the SED, and disk flaring angle, based on
the ratio of the near- to mid-IR excess.  The latter results argue for
coagulation and grain settling towards the mid-plane as the explanation for the
observed bi-modal distribution of SEDs, consistent with our findings here for
lower mass solar-type stars. 

We suspect that the correlation from IR
spectroscopy of the derived grain size and the disk flaring has been missed so
far for the HAEBE systems are due to the at least one order
of magnitude higher stellar luminosity of A and B type stars compared to the
later-type stars studied here. Observations of HAEBE systems
will, therefore,  probe a much larger region of the circumstellar disk 
than the observations of TTS or brown dwarf systems for any given wavelength,
by at least a factor 2 to 3. Given that coagulation
and settling are a function of density and thus radius, gradients in the observed
grain size and disk structure are to be expected. Probing a much larger region of the 
disk, therefore, could lead to the loss of any clear correlations In the HAEBE systems.

\subsection{The formation of crystalline silicates}
All systems studied in this paper show emission from crystalline silicates,
further evidence for dust processing within the protoplanetary disks as
interstellar dust, the material present at the formation time of the disks,
shows no evidence for crystalline silicates.  The results of the spectral
analysis of the crystalline silicate emission can be summarized as follows:
\begin{enumerate}
\item All observed targets show emission bands of crystalline silicates at both
higher ($\sim$300 - 500~K) temperatures, probed in the 10~$\mu$m spectral window,
and lower ($\sim$100~K) temperatures, probed in the 30~$\mu$m spectral region.
\item The observed emission bands are consistent with emission from the pure magnesium
silicates forsterite and enstatite, and silica. No conclusive evidence for other
species has been found. 
\item The average grain size of the crystals is much
smaller than that of the amorphous silicates. While amorphous silicates
requires grain sizes up to 6~$\mu$m, the crystalline emission is
more consistent with a size of $\sim$0.1 to 1~$\mu$m. 
\item we find no conclusive evidence for a correlation between the mass fraction of the crystalline silicates and the grain size of the amorphous silicates. 
\item The average grain size of the enstatite grains as derived from the
modelling of the 10~$\mu$m spectral region is systematically larger than that of
the forsterite grains. 
\item The strength of the forsterite emission features in the 10~$\mu$m
wavelength region and those at the longer wavelengths seem to be correlated. This could
be interpreted as evidence that the crystalline mass fraction is increased from
ISM values to the observed mass fractions of about $\sim$5\%, throughout the
entire inner $\sim$15~AU of the disk. 
\item We find a change in the relative abundance of the different crystalline
silicates from the very inner ($\sim$1~AU) warm dust population, which is
dominated by enstatite, compared to the population at the colder outer disk
regions at larger radii ($\sim$5-15~AU), dominated by forsterite. This change in
relative abundances points towards a radial dependence of either the formation
mechanism of the crystalline silicates, or the (non) equilibrium conditions
under which they formed, and also argues against substantial radial mixing of
processed material from the inner to the outer parts of the disk (see paragraphs
below for a detailed discussion).
\end{enumerate}

How can we interpret the above results for the crystalline silicates in a
consistent picture of the physical processes occurring in a protoplanetary disk? As
all our targets show similar crystalline mass fractions and compositions, these
physical processes can be expected to occur in all protoplanetary disks. The
formation of forsterite and enstatite requires temperatures above $\sim$1000~K
\citep[e.g.][]{hallenbeck2000,fabian2000}. Such high temperatures are reached near the central
star where dust grains can be readily heated by the stellar radiation field.
In an early active disk phase, the accretion energy can
provide an additional heat source for the dust in the mid-plane regions of the inner disk
\citep[e.g.][]{bell2000}. At these high temperatures, crystalline silicates can
be formed either through annealing  of the amorphous silicates \citep[see][for
an overview]{wooden2005} or by gas-phase condensation/annealing and gas-solid reactions
in a cooling gas \citep{davis2003, petaev2005}. Our observations show, however,
that a substantial fraction of the crystalline silicates have temperatures
($\sim$100~K) far below the required crystallization temperature. This implies
that either large scale radial mixing has to occur, transporting the crystalline
silicates from the hot inner parts to the cooler outer disk
\citep[e.g.][]{dominique2002}, or that an additional localised and/or transient
heating mechanism is operating in the cooler regions of the disk, at larger
radial distances from the central star. Such additional heating mechanism might
be provided by shock-waves in which gas and dust grains are heated to the
required temperatures of above a 1000~K \citep{harker2002,Desch2005}.

The question is now if our results can distinguish between different 
formation mechanisms and/or locations for the crystalline silicates.
The presence of forsterite and enstatite, respectively the magnesium 
rich end members of the olivine and
pyroxene solid-solution series, with no evidence for crystalline silicates
containing substantial amounts of iron, suggest they formed as high-temperature,
gas-phase condensates \citep[][]{davis2003, gail2004}. The presence of
high-temperature condensates at lower temperatures indicates that complete
equilibrium condensation is not taking place in protoplanetary disks, else these
species would have been transformed to other ferromagnesian silicates during
cooling. Though the exact order in which the different dust species condense out
from a cooling gas depends on the gas pressure and isolation of the condensed
dust grains from the surrounding gas \citep{petaev2005}, most models predict
forsterite to be condensed out first, followed by the formation of enstatite at
slightly lower temperatures through solid-gas reactions between forsterite and
SiO$_2$ gas. Iron will condense out slightly before or after forsterite as
metallic iron grains rather than being incorporated into silicates. The slightly
larger grain size of enstatite compared to forsterite can be the result from the
gas-solid reactions and the broader temperature range (and thus a longer
formation window in a slowly cooling gas) under which enstatite can be formed
and is stable.

An alternative formation mechanism for forsterite and enstatite is annealing of
amorphous silicates without evaporating/recondensating the parent grain. 
The annealing process critically depends on the structure
of the amorphous material and the oxygen partial pressure of the surrounding gas
(i.e. reducing or oxidizing conditions). To prevent the formation of iron
containing olivine and pyroxene minerals, the amorphous grains have to be
annealed under reducing conditions. \cite{davoisne2006} show that carbon
combustion locking up the oxygen in the form of CO rather then in FeO, will
cause the iron the precipitate as metallic particles, and not get incorporated
into a silicate. \cite{thompson2002} showed that forsterite and silica,
independent of the stoichiometry of the amorphous material, will always form,
but that the formation of enstatite follows through solid-solid reactions 
between forsterite and silica. The effectiveness and speed of this reaction
critically depends on the structure of the amorphous grains. In highly porous
grains, like those formed from smokes, the contact surfaces between the
individual sub-units will be small and, therefore, reactions will be slow and require
substantial structural modification (like melting) of the amorphous grain. For
compact glassy amorphous silicates, the formation of enstatite can proceed
immediately.  Similar as with direct condensation, this secondary formation of
enstatite through forsterite alteration could also explain why we observe a
slightly larger grain size for enstatite compared to forsterite.

Apart from the above considerations two additional observational constraints can
be imposed to distinguish between annealing or condensations as the main
formation mechanism for crystalline silicates, namely summary points (3) and
(4). The implication of these two observations is that the simultaneous
presence of both large amorphous grains and small crystalline grains is
\emph{not} caused by a preferential transformation of small amorphous grains
into crystalline silicates, leaving the larger amorphous grains. Or with other words,
the observed change in grain size towards larger grains is caused by grain growth
rather than the removal of the smaller amorphous grain population trough crystallization
in a distribution of grain sizes. 
If the latter would have been the case, a correlation between amorphous grain size and crystalline mass fraction would have been observed. This together with the fact that no
large crystalline grains are observed implies that either the crystals form as
gas-phase condensates, the large amorphous aggregates are disrupted before
annealing, or that the main crystallization process occurred before coagulation.

Gas-phase condensation can naturally explain any difference in grain size
between the amorphous and crystalline grains. In case of an
evaporation/condensation zone in the disk, where first the amorphous grains are
evaporated after which crystalline species can condense out if the gas cools, no
relation between the original grain size distribution of the amorphous dust and
that of the crystalline grains will exist. Annealing of large amorphous grains,
on the other hand, will lead to the formation of large crystals. It is,
therefore, necessary to disrupt the larger aggregates before they get annealed.
A gradual heating of dust grains as they approach the hot inner disk is unlikely
to fulfil this requirement. Annealing by shock waves could be a more plausible
scenario, as it could also provide the mechanism to disrupt the larger amorphous
aggregates. This formation mechanism would be similar as the formation scenario 
of chondrules \citep[e.g.][]{Desch2005}.

In case of an early formation time of the
crystalline silicates, that is before coagulation and during the high accretion
phases, viscous dissipation can efficiently heat the disk up to 
several AU from the central star \citep[e.g][]{bell2000}. An early
formation time would imply that the crystalline silicate content of
protoplanetary disks would remain fixed during their further evolution. This is
consistent with previous studies who reported no correlations between the
processing of the silicates and systemic ages
\citep[e.g.][]{Roy2005,Kessler-Silacci2006, Kessler-Silacci2005}, though any
correlation might also be lost due to the large uncertainties in age
determinations. On the other hand, all systems studied in this paper show
crystalline silicates. Studies of on average younger samples show that a
substantial fraction of disks show no crystalline silicates \citep[e.g.
50~\%][]{Kessler-Silacci2006}. This would argue in favour of crystallinity
increasing with time. Further, recent studies of FU Orionis objects show no sign of
crystalline silicates, independent of whether the silicate emission originates
in disks or envelopes \citep[][]{Quanz2006, Green2006,Quanz2007}.
If the notion is correct that the FU~Orionis type objects
represent the early high accretion stages of TTS evolution, crystalline
silicates should be present if the bulk of the crystalline silicates would form
during the early evolution of the disks. Interestingly, \cite{Quanz2006} show
that FU~Orionis already shows signs of grain growth. If coagulation sets in this
early, an alternative scenario for the formation of crystalline silicates is
required.

Note that there is no difference in coagulation behaviour between crystalline
and amorphous silicates. The difference in grain size between the crystalline
and amorphous silicates can be interpreted in terms of the optical properties of
larger composite aggregates. Suppose a population of small, predominantly
amorphous dust grains coagulates into larger aggregates. These larger aggregates
will obviously consists mainly of amorphous material but with few, from each
other isolated crystalline grains which will therefore react with the radiation
field as separate entities. The optical properties of such composite aggregates
will likely resemble that of the combined properties of a larger amorphous grain
with smaller crystalline particles, similar as is observed in cometary spectra
and IDPs. 

The two remaining summary points, (6) and most importantly (7), to be discussed here, 
place further constraints on the formation mechanism of the crystalline silicates and
on the location in the disk where the crystalline silicates are forming. On the
one hand our data suggests a general increase in the crystalline mass fraction in the
inner $\sim$15~AU of the disks, on the other hand we observe a gradient in the
crystalline dust composition within this inner region. The changing forsterite
over enstatite mass ratio points to differences in the condensation or annealing
conditions. As discussed in the previous paragraphs, the formation of enstatite
follows after the formation of forsterite. The absence or much lower enstatite
mass fraction derived from the longer wavelength bands, probing the cooler dust
at about 5-15~AU compared to the mass ratio derived from the 10~$\mu$m
wavelength region, shows that much less or even none of the forsterite is converted
into enstatite at larger radial distances from the central star. In the case of
gas-phase condensation as the formation mechanism for the crystalline silicates,
this implies non-equilibrium condensation conditions at larger disk radii, in
contrast to the very inner disk regions. If forsterite forms locally at
$\sim$5-15~AU in the disk it has to be isolated from the surrounding SiO gas
before they react to form enstatite. This could be achieved by
coagulation, or by rapid cooling of the surrounding gas, both inhibiting the
gas-solid reactions leading to the formation of enstatite.  Alternatively, in
case forsterite is formed through annealing of ISM material, the low enstatite
to forsterite ratio implies a rather porous structure of the amorphous silicates
and brief heating events, like shocks, preventing the formation of enstatite.

The high temperature region in the inner disk near the central star, is the
natural location where crystallization must occur. 
Assuming that the inner disk region is the sole region where crystalline silicates 
are formed, the presence of crystalline silicates at low temperatures
implies that efficient radial mixing of the processed dust from the inner disk
to larger radii must occur. At the early, high accretion phases of the disk, the
transport of angular momentum is expected to efficiently mix material from the
hot inner to the cold outer parts of the disk
\citep[][]{dominique2002,dullemond2006}. Though at a first glance this scenario
would be able to explain our observations, a strong argument against it is the radial
dependency of the crystalline silicate composition: If the crystalline silicates
originates from a single formation region before being distributed through the
disk, the composition of the processed material would be constant with radius.
Indeed, detailed disk models by \cite{gail2004}, taking into account radial
mixing, predicting the dust composition as a function of radius, show exactly
this counter argument.  While  \cite{gail2004} predict the composition of the
crystalline silicates in inner regions of the disk to be dominated by enstatite,
consistent with our observations, the radial mixing model will also predict a
large mass fraction of enstatite at larger radii, in contradiction with our
results. The only way to save the radial mixing scenario is if during the active
mixing phase, non-equilibrium conditions prevailed in the inner parts of the
disk, forming only forsterite. This might be achieved by very rapid outward
transportation of the formed forsterite preventing prolonged heating of the
grains or substantial gas-solid reactions which could lead to the formation of
enstatite. During the passive, low accretion disk phases, the phase were we are
observing our TTS sample, were radial mixing is not expected to be efficient,
equilibrium conditions might be reached again, resulting in the observed large
fraction of enstatite in the inner disks. At present, however, there is no
theoretical foundation for this scenario.
 
The alternative model to the radial mixing scenario is the in-situ
formation of crystalline silicates at radii of about 5-15~AU. As at these distances stellar radiation can not heat the dust to sufficiently high temperatures for
crystallization to occur, an alternative heating mechanism is required. A
promising model is the local heating of dust and gas by shock waves
\citep[][]{harker2002, Desch2005}.  For typical densities and shock speeds,
small dust grains should be efficiently heated to high temperatures up to disk
radii of about 10~AU. A possible source for the shocks could be gravitational
instabilities or planetary mass companions within the disk. Shock heating could
provide sufficiently brief heating events such that equilibrium conditions or
prolonged annealing will not occur, preventing the formation of enstatite at
larger radii. In the very inner ($\sim$1~AU) disk, at higher densities,  apart
from eating by the central star, cooling can be expected to be longer such that
equilibrium conditions could be reached. 
The suggested overall increase of the forsterite
mass fraction in the  inner $\sim$15~AU of the disks, might reflect that spiral
waves could affect the entire disk region, causing shock to occur over a large
range of radii or simply that the forsterite, produced at 5-15~AU from the
central star, is accreted inwards to smaller radii. 
This shock-wave scenario
could explain why some systems do not show any evidence for the presence of
crystalline silicates at the longest or even all wavelengths
\citep[e.g.][]{Kessler-Silacci2006}: these systems might not (yet) have produced
disk instabilities or sufficiently large planetary sized objects to produce
shock-waves.

Ideally, the spatial distribution of the different silicate species, should be
directly determined by spatially resolved spectroscopy. Due to observational
difficulties, this is at present limited to the much brighter HAEBE and the most
nearby TTS systems. Using mid-IR interferometric observations obtained with the
VLTI/MIDI instrument, \cite{roy2004} could spatially separate contribution to
the IR emission coming from the inner $\sim$2~AU of the disk, and that coming
from larger radii, for a sample of HAEBE systems. These observations clearly
showed that the crystalline silicates in the observed HAEBE systems are mainly
concentrated in the inner, high temperature, parts of the disk, and have a
composition consistent with a formation  under (near) equilibrium conditions.
However, before concluding that all crystalline silicates form in the inner parts
of the disk, one has to realize that of the by \cite{roy2004} studied systems,
show no or substantially weaker crystalline silicate bands at the longer
(20-30~$\mu$m) wavelengths in their Spitzer and ISO spectra \citep[][Bouwman et
al, in prep.]{HerbigOverview}. This suggest that these HAEBE systems, in
contrast to the in disks in our TTS sample, did not witnessed substantial radial
mixing or in-situ formation of crystalline silicates at larger radii, with only the very
inner regions of the HAEBE disks containing observable quantities of  
crystalline silicates. 

\subsection{PAH molecules in disks surrounding T~Tauri stars.}
Though the main focus of this paper is the silicate dust processing we report
the tentative detection of emission bands from polycyclic aromatic hydrocarbon
molecules. Emission bands of PAH molecules are commonly found in environments
with a strong UV flux like the circumstellar disks around HAEBE stars. As PAH
molecules can contain a substantial fraction of the available carbon, and are
spectroscopically easier detectable as carbon in larger grains, they can provide
us with important clues concerning the carbon chemistry of circumstellar
environments. Characterizing the PAH population is also important for understanding the 
observed gas temperatures in disks, as the stochastically heated molecules strongly influence the temperature through photo-electric heating. Since the gas temperature determines the pressure scaleheight of the disk, the presence of PAH molecules can, therefore, have a substantial 
influence on the disk geometry. Five out of the seven TTS systems of our sample show a band at 8.2~$\mu$m we identify with emission from PAH molecules. 
This is the first time this band is observed in low-mass pre-main-sequence systems. 
The relative high fraction of systems in our sample showing
evidence for PAH molecules seems to be contradicting other studies such as the
Core to Disks Spitzer legacy study by \cite[][]{Geers2006}, which found no PAH
emission around sources with spectral types later than G8. However, many studies
are based on the analysis of the 6.2 and 11.3~$\mu$m PAH bands. As already
noticed by \cite{peeters2002}, these latter two bands seem to be suppressed in
the PAH population producing the 8.2 $\mu$m feature. Just based on the 6.2 and
11.3~$\mu$m features, we would only have detected PAH emission in the HD~143006
system, a G6/8-type star, which would be consistent with the conclusions by
\cite[][]{Geers2006}. Also, as the 8.2 $\mu$m feature coincides with the strong
amorphous silicate band, one can only make a firm conclusion concerning its
presence after carefully modelling the silicate emission bands. The fact that we
observe a PAH band at 8.2~$\mu$m rather than the commonly observed band position
between 7.7 to 7.9$\mu$m suggests a fundamental difference in carbon chemistry
between our sample and the more luminous HAEBE and F- and early G-type TTS
systems. At this point we can only speculate that the difference in carbon
chemistry could be linked to differences in X-ray fluxes or stellar wind
properties of the intermediate-mass systems compared to those of low-mass
pre-main-sequence stars. If so this could also influence the possibility and
chemical path by which more complex pre-biotic organic molecules can be formed
in protoplanetary disks.

\appendix

\section {HD~143006}

HD~143006 is a well-studied \citep[e.g.][]{Garcia-Lario1990, Walker1988, 
sylvester1996, Coulson1998, Natta2004, Dent2005}
classical T Tauri star of spectral type G6/8 which appears
to be member of the Upper Scorpius OB association \citep[][]{Mamajek2004}.
It displays many T~Tauri characteristics including infrared to millimeter
excess \citep[][]{Odenwald1986}, 
optical emission lines \citep[][]{Stephenson1986}, 
and coronal activity revealed through x-rays \citep[][]{Sciortino1998}.
It is thus an ``old" accretion disk system, with an age of 3-5~Myr.

\section {RX~J1612.6-1859A}

RX~J1612.6-1859A and RX~J1612.6-1859B are the names assigned in the nomenclature
of \cite{Martin1998} for a pair of emission line sources in the $\rho$~Oph
vicinity.  The former object is an M0 star also known as 2MASS~J16123916-1859284
or GSC~06209-01312 that has been studied only by optical spectroscopy
\citep{Martin1998,The1964}. The latter object, by contrast, is a well-known T
Tauri star, also known as ScoPMS~52 \citep[][]{Walter1994} or Wa~Oph~3
\citep[][]{Walter1986}. This K0 star was, in fact, the target of the FEPS
observations and the source discussed in our companion paper
\citep{Silverstone2006} as not having any excess emission out to 8 $\mu$m (and
indeed not out to 24 $\mu$m; Carpenter et al., in preparation).  For our IRS
observations of this source, however, the IRS peak-up routine centered on to the
mid-IR bright object RX~J1612.6-1859A in the slit rather than RX~J1612.6-1859B
20" away.  The observations discussed herein are thus of RX~J1612.6-1859A, an M0
emission line object.

\acknowledgements  

FEPS gratefully acknowledges support from NASA through JPL
grants 1224768, 12224634, and 1224566.
We would like to thank Dan Watson and Pat Morris for helpful
discussions regarding data reduction, Deborah Padget
and Tim Brooke for assistance with the observing plan,
and Betty Stobie for assistance with software development,
as well as the rest of the FEPS team, the
IRS instrument team, and colleagues at the Spitzer Science
Center for their support.  MRM is also supported through NASA's
Astrobiology Institute. JB and ThH acknowledge support from the
EU Human Potential Network contract No. HPRN-CT-2002000308.

\bibliographystyle{aa}
\bibliography{reference_list,publication_list}

\clearpage
\begin{deluxetable}{llccccccccc}
\tabletypesize{\scriptsize}
\rotate
\tablecaption{Astrophysical parameters of program stars.\label{tbl:stars}}
\tablewidth{0pt}
\tablecolumns{11}
\tablehead{
\colhead{ID\#} & \colhead{Name} & \colhead{$\alpha$} & \colhead{$\delta$} & \colhead{$d$} & \colhead{log(Age)} & \colhead{Sp. Type} & \colhead{Ref.} & \colhead{$T_{\rm eff}$} & \colhead{$A_V$} & \colhead{L$_star$}\\
\colhead{} & \colhead{}  &   \colhead{(2000)} & \colhead{(2000)} & \colhead{[pc]} & \colhead{[yr]} & \colhead{} & \colhead{[K]} & \colhead{} & \colhead{[mag]}  & \colhead{[L$_\odot$]}
}
\startdata
0 &RX J1842.9-3532       &18:42:57.98  &-35:32:42.73 & 145 & 6.63  (4 Myr) & K2    & 2,3,1 & 4995  & 1.03 & 1.0\\
1 &RX J1852.3-3700       &18:52:17.30  &-37:00:11.93 & 145 & 6.5-7 (4 Myr) & K3    & 2,3,1 & 4759  & 0.92 & 0.6\\
2 &HD 143006             &15:58:36.92  &-22:57:15.35 & 145 & 6.70  (5 Myr) & G6/8  & 7,8,9 & 5884  & 1.63 & 2.5\\
3 &RX J1612.6-1859A      &16:12:39.18  &-18:59:28.0  & 145 & 6.70  (3 Myr) & M0    & 11    & 3800  & 1.80 & 0.5\\
  &(2MASS J16123916-1859284)&             &            &     &               &       &       &       &      &    \\
4 &1RXS J132207.2-693812 &13:22:07.53  &-69:38:12.18 &  86 & 7.23  (17 Myr) & K1IVe & 10    & 5228  & 1.22 & 1.3\\
  &(PDS 66)              &             &            &     &               &       &       &       &      &    \\
5 &RX J1111.7-7620       &11:11:46.32  &-76:20:09.21 & 163 & 6.69  (5 Myr) & K1	   & 5,3,4 & 4621  & 1.30 & 1.6\\
6 &1RXS J161410.6-230542 &16:14:11.08  &-23:05:36.26 & 145 & 6.70  (5 Myr) & K0    & 7,8,6 & 4963  & 1.48 & 3.2\\
  &([PZ99] J161411.0-230536)&             &            &     &               &       &       &       &      &    \\
\tableline
\enddata

\tablenotetext{~}{
References for distance, age, spectral type -- 
(1) Neuhauser et al. 2000;
(2) Neuhauser \& Forbrich 2007;
(3) Hillenbrand et al. 2007, in preparation;
(4) Alcala et al. 1995;
(5) Luhman 2007;
(6) Preibisch et al. 1998;
(7) de Zeeuw et al. 1999;
(8) Preibisch et al. 2002;
(9) Houck \& Smith-Moore 1988;
(10) Mamajek et al. 2002;
(11) Martin et al. 1998.
}
\tablenotetext{~}{
Notes for temperature, extinction, luminosity --
Spectral types are from optical spectroscopy as cited above.
B-V and V-K colors are used in conjunction with spectral types
to estimate effective temperatures and visual extinction values;
see Carpenter et al. 2007, in preparation, for details.
In the case of RX J1612.6-1859A J-H and H-K colors are used and
luminosity is computed from J-band bolometric correction.
}
\end{deluxetable}

\clearpage

\begin{deluxetable}{llllll}
\tabletypesize{\small}
\tablewidth{0pt}
\tablecolumns{6}
\tablecaption{Overview of dust species used. For each component we specify
 its lattice structure, chemical composition, shape and reference to the 
 laboratory measurements of the optical constants. For the homogeneous
 spheres we used Mie theory to calculate the opacities. For the inhomogeneous
 spheres, we used the distribution of hollow spheres \citep[DHS;][]{min2005}, to simulate
 grain deviating from perfect symmetry. \label{tbl-species} }
\tablehead{\colhead{\#}&\colhead{Species}&\colhead{state}&\colhead{Chemical}&\colhead{Shape}&\colhead{Ref} \\
 \colhead{}& \colhead{}&\colhead{} & \colhead{ formula}  &  \colhead{}  & \colhead{} } 

\startdata
1 & Amorphous silicate  & A  &  MgFeSiO$_{4}$       & Homogeneous   & (1)  \\
  & (Olivine stoichiometry)&  &                     &                &  \\
2 & Amorphous silicate  & A  &  MgFeSi$_{2}$O$_{6}$ & Homogeneous   & (1)  \\
 & (Pyroxene stoichiometry)&  &                     &                &  \\
3 & Forsterite   & C  &  Mg$_{2}$SiO$_{4}$   & Inhomogeneous & (2)   \\
4 & Clino Enstatite    & C  &  MgSiO$_{3}$         & Inhomogeneous & (3) \\
5 & Silica       & A  &  SiO$_{2}$           & Inhomogeneous & (4)  \\
\tableline
\enddata
\tablerefs{(1)\citet{dorschner1995}; (2)\citet{servoin1973}; (3)\citet{jaeger1998}; (4)\citet{Henning1997};}
\end{deluxetable}

\clearpage
\pagestyle{empty}

\begin{deluxetable}{@{} rlllllllllllllll @{}}
\setlength{\tabcolsep}{0.02in} 
\tabletypesize{\tiny}
\rotate
\tablecaption{ The best fit values of the parameters in our compositional fits to the 8 to 13~$\mu$m spectral region.The abundances of the various dust species are given
as a percentage of the total dust mass, \emph{excluding} the dust responsible
for the continuum emission. Also listed is the mass averaged grain size of the different dust species.
If a species was not found, or unconstrained by the spectra, this is indicated by a -~symbol.
The PAH and continuum flux contributions are listed as percentages of the total integrated 
flux over the 10\,$\mu$m region, contained in these components. 
These are measures for the relative flux contributions, but cannot be interpreted as relative dust masses.
The resulting best fit model spectra are plotted in Fig.~\ref{fig:fit} (light grey lines).
\label{tab:t2_fs:fits} }
\tablewidth{0pt}
\tablecolumns{16}


\tablehead{
\colhead{\#} & \colhead{$\chi^2$} & \colhead{$T_{c}$\,[K]} & \colhead{$T_{dust}$\,[K]} &
\colhead{Cont.}  & \multicolumn{2}{c}{Amorph. Olivine}  & \multicolumn{2}{c}{Amorph. Pyroxene}  &
\multicolumn{2}{c}{Forsterite}  & \multicolumn{2}{c}{Enstatite}  &
\multicolumn{2}{c}{Silica}  &\colhead{PAH} \\
  \colhead{}  &   \colhead{}    & \colhead{}     &    \colhead{}        & \colhead{contr.} &  
\colhead{mass } &\colhead{ $<a>$ }  & \colhead{mass}   & \colhead{$<a>$}  & \colhead{mass } &
\colhead{$<a>$} &\colhead{mass}   & \colhead{$<a>$ }  & \colhead{mass } & \colhead{$<a>$} &\colhead{contr.}
}

\startdata
\multicolumn{16}{l}{\bf 8-13$\mu$m spectral region} \\
0 & 11.1$\pm$0.5 & 1500.0$_{-0.0}^{+0.0}$ & 324.6$_{-5.6}^{+6.9}$ & 0.5214$_{-0.0009}^{+0.0009}$ & 
0.7902$_{-0.0063}^{+0.0063}$ & 1.16$_{-0.09}^{+0.09}$ & 
0.1787$_{-0.0072}^{+0.0071}$ & 1.12$_{-0.08}^{+0.09}$ & 
0.0146$_{-0.0004}^{+0.0004}$ & 0.1$_{-0.0}^{+0.0}$ & 
0.0165$_{-0.0015}^{+0.0016}$ & 1.20$_{-0.09}^{+0.08}$ & 
  -    &   -  & 
0.0066$_{-0.0004}^{+0.0004}$\\
1 & 3.4$\pm$0.3 & 1500.0$_{-0.0}^{+0.0}$ & 257.4$_{-6.2}^{+3.9}$ & 0.6361$_{-0.0025}^{+0.0027}$ & 
0.9872$_{-0.0058}^{+0.0042}$ & 3.08$_{-0.13}^{+0.07}$ & 
0.0016$_{-0.0014}^{+0.0064}$ & 0.1$_{-0.0}^{+0.0}$ & 
0.0016$_{-0.0009}^{+0.0011}$ & 0.1$_{-0.0}^{+0.0}$ & 
0.0091$_{-0.0038}^{+0.0036}$ & 1.41$_{-0.32}^{+0.08}$ & 
0.0006$_{-0.0006}^{+0.0030}$ & 1.5$_{-0.0}^{+0.0}$ & 
0.0082$_{-0.0017}^{+0.0015}$\\
2 & 17.2$\pm$0.7 & 1459.0$_{-13.5}^{+17.1}$ & 296.2$_{-2.6}^{+5.3}$ & 0.6255$_{-0.0013}^{+0.0011}$ & 
0.4225$_{-0.0053}^{+0.0120}$ & 0.13$_{-0.03}^{+0.47}$ & 
0.5296$_{-0.0137}^{+0.0047}$ & 5.30$_{-0.05}^{+0.06}$ & 
0.0149$_{-0.0003}^{+0.0003}$ & 0.1$_{-0.0}^{+0.0}$ & 
0.0330$_{-0.0011}^{+0.0010}$ & 1.496$_{-0.02}^{+0.004}$ & 
   -        &            - & 
0.0246$_{-0.0003}^{+0.0003}$\\
3 & 8.8$\pm$0.7 & 1030.6$_{-94.7}^{+77.5}$ & 369.9$_{-44.0}^{+66.4}$ & 0.6792$_{-0.0210}^{+0.0263}$ & 
0.2933$_{-0.2375}^{+0.2511}$ & 6.00$_{-0.009}^{+0.0}$ & 
0.6452$_{-0.2529}^{+0.2366}$ & 4.85$_{-1.94}^{+0.34}$ & 
0.0141$_{-0.0018}^{+0.0044}$ & 0.15$_{-0.04}^{+0.24}$ & 
0.0342$_{-0.0054}^{+0.0061}$ & 1.13$_{-0.13}^{+0.13}$ & 
0.0133$_{-0.0027}^{+0.0040}$ & 0.37$_{-0.22}^{+0.27}$ & 
0.0103$_{-0.0021}^{+0.0031}$\\
4 & 11.1$\pm$0.5 & 778.8$_{-14.6}^{+2.7}$ & 183.8$_{-14.8}^{+1.2}$ & 0.6054$_{-0.0011}^{+0.0031}$ & 
0.6369$_{-0.0412}^{+0.0367}$ & 5.57$_{-0.05}^{+0.08}$ & 
0.3279$_{-0.0370}^{+0.0404}$ & 2.27$_{-0.75}^{+0.45}$ & 
0.0114$_{-0.0005}^{+0.0003}$ & 0.1$_{-0.0}^{+0.0}$ & 
0.0238$_{-0.0026}^{+0.0013}$ & 0.97$_{-0.07}^{+0.04}$ & 
  -     &  -  & 
0.0126$_{-0.0004}^{+0.0007}$\\
5 & 4.0$\pm$0.4 & 1500.0$_{-0.0}^{+0.0}$ & 236.6$_{-7.2}^{+6.5}$ & 0.8544$_{-0.0010}^{+0.0011}$ & 
0.8236$_{-0.0160}^{+0.0138}$ & 6.0$_{-0.0}^{+0.0}$ & 
0.1060$_{-0.0133}^{+0.0153}$ & 0.1$_{-0.0}^{+0.0}$ & 
0.0272$_{-0.0023}^{+0.0025}$ & 0.97$_{-0.06}^{+0.05}$ & 
0.0349$_{-0.0022}^{+0.0023}$ & 1.39$_{-0.08}^{+0.07}$ & 
0.0083$_{-0.0015}^{+0.0019}$ & 0.35$_{-0.21}^{+0.27}$ & 
0.0022$_{-0.0005}^{+0.0005}$\\
6 & 6.9$\pm$0.4 & 617.0$_{-7.1}^{+13.2}$ & 305.1$_{-5.8}^{+5.6}$ & 0.8206$_{-0.0096}^{+0.0055}$ & 
0.5862$_{-0.1153}^{+0.0810}$ & 6.0$_{-0.0}^{+0.0}$ & 
0.3728$_{-0.0802}^{+0.1200}$ & 5.999$_{-0.016}^{+0.001}$ & 
0.0058$_{-0.0005}^{+0.0005}$ & 0.1$_{-0.0}^{+0.0}$ & 
0.0249$_{-0.0024}^{+0.0022}$ & 1.21$_{-0.09}^{+0.08}$ & 
0.0104$_{-0.0009}^{+0.0009}$ & 0.1$_{-0.0}^{+0.0}$ & 
0.0066$_{-0.0011}^{+0.0008}$\\
\tableline
\multicolumn{16}{l}{\bf 17-26$\mu$m spectral region} \\
0 & 28.3$\pm$1.4 & 46.3$_{-1.4}^{+3.8}$ & 118.0$_{-0.0}^{+0.0}$ & 0.0321$_{-0.0018}^{+0.0022}$ & 
0.9592$_{-0.0071}^{+0.0066}$ & 1.5$_{-0.0}^{+0.0}$ & 
0.0161$_{-0.0069}^{+0.0073}$ & 0.80$_{-0.67}^{+0.67}$ & 
0.0192$_{-0.0005}^{+0.0009}$ & 0.14$_{-0.04}^{+0.11}$ & 
    -     &       -        & 
0.0055$_{0.0005}^{+0.0005}$ & 1.5$_{0.0}^{+0.0}$ & 
     -             \\
1 & 7.0$\pm$0.7 & 105.3$_{-0.6}^{+30.1}$ & 84.95$_{-0.0}^{+0.05}$ & 0.2529$_{-0.0289}^{+0.0186}$ & 
0.8650$_{-0.0119}^{+0.0136}$ & 1.61$_{-0.09}^{+0.14}$ & 
0.1141$_{-0.0136}^{+0.0122}$ & 1.5$_{-0.0}^{+0.0}$ & 
0.0099$_{-0.0013}^{+0.0011}$ & 0.72$_{-0.24}^{+0.21}$ & 
0.0007$_{-0.0006}^{+0.0015}$ & 0.1$_{-0.0}^{+0.0}$ & 
0.0104$_{-0.0007}^{+0.0007}$ & 1.5$_{-0.0}^{+0.0}$ & 
        -       \\
2 & 18.8$\pm$1.0 & 33.0$_{-3.3}^{+5.4}$ & 112.7$_{-3.9}^{+0.3}$ & 0.0096$_{-0.0024}^{+0.0047}$ & 
0.7384$_{-0.0195}^{+0.1520}$ & 1.5$_{-0.0}^{+0.0}$ & 
0.2459$_{-0.1474}^{+0.0201}$ & 5.84$_{-0.08}^{+0.10}$ & 
0.0100$_{-0.0004}^{+0.0005}$ & 0.1$_{-0.0}^{+0.0}$ & 
     -       &      -       & 
0.0057$_{0.0004}^{+0.0008}$ & 1.5$_{0.0}^{+0.0}$ & 
 -             \\
3 & 5.9$\pm$0.6 & 18.2$_{-3.2}^{+6.0}$ & 197.7$_{-6.4}^{+7.9}$ & 0.0013$_{-0.0006}^{+0.0010}$ & 
0.1214$_{-0.0353}^{+0.0348}$ & 1.5$_{-0.0}^{+0.0}$ & 
0.8628$_{-0.0348}^{+0.0369}$ & 5.67$_{-0.03}^{+0.04}$ & 
0.0130$_{-0.0007}^{+0.0007}$ & 0.1$_{-0.0}^{+0.0}$ & 
    -      &     -           & 
0.0029$_{-0.0008}^{+0.0008}$ & 0.71$_{-0.55}^{+0.71}$ & 
      -           \\
4 & 13.0$\pm$0.8 & 106.0$_{-7.4}^{+5.4}$ & 152.6$_{-3.9}^{+2.6}$ & 0.1748$_{-0.0388}^{+0.0513}$ & 
0.3846$_{-0.0594}^{+0.0912}$ & 1.5$_{-0.0}^{+0.0}$ & 
0.5881$_{-0.0967}^{+0.0614}$ & 4.99$_{-0.39}^{+0.18}$ & 
0.0226$_{-0.0023}^{+0.0039}$ & 0.1$_{-0.0}^{+0.0}$ & 
0.0001$_{-0.0001}^{+0.0014}$ & 0.1$_{-0.0}^{+0.0}$ & 
0.0047$_{-0.0009}^{+0.0010}$ & 1.13$_{-0.68}^{+0.34}$ & 
     -                 \\
5 & 5.3$\pm$0.5 & 177.0$_{-2.6}^{+3.9}$ & 294.5$_{-31.3}^{+34.6}$ & 0.7317$_{-0.0104}^{+0.0120}$ & 
0.0328$_{-0.0311}^{+0.0784}$ & 6.0$_{-0.0}^{+0.0}$ & 
0.8255$_{-0.0717}^{+0.0368}$ & 1.5$_{-0.0}^{+0.0}$ & 
0.0996$_{-0.0071}^{+0.0090}$ & 1.26$_{-0.10}^{+0.10}$ & 
0.0206$_{-0.0077}^{+0.0088}$ & 0.1$_{-0.0}^{+0.0}$ & 
0.0215$_{-0.0051}^{+0.0052}$ & 0.15$_{0.05}^{+0.56}$ & 
      -             \\
6 & 3.9$\pm$0.5 & 340.0$_{-110.6}^{+9.7}$ & 155.1$_{-13.1}^{+71.6}$ & 0.6692$_{-0.1003}^{+0.0308}$ & 
   -            &     -          & 
0.9643$_{-0.0255}^{+0.0060}$ & 1.43$_{-0.83}^{+0.85}$ & 
0.0321$_{-0.0051}^{+0.0157}$ & 0.52$_{-0.29}^{+0.23}$ & 
0.0025$_{-0.0021}^{+0.0040}$ & 0.1$_{-0.0}^{+0.0}$ & 
0.0012$_{-0.0011}^{+0.0106}$ & 0.1$_{-0.0}^{+0.0}$ & 
          -               \\
\tableline
\multicolumn{16}{l}{\bf 26.5-35.5$\mu$m spectral region} \\
0 & 15.3$\pm$2.0 & 25.7$_{-1.6}^{+4.9}$ & 96.7$_{-2.7}^{+5.2}$ & 0.0726$_{-0.0115}^{+0.0282}$ & 
0.0458$_{-0.0396}^{+0.1059}$ & 2.95$_{-2.71}^{+2.75}$ & 
0.9341$_{-0.1055}^{+0.0392}$ & 0.66$_{-0.45}^{+0.65}$ & 
0.0165$_{-0.0009}^{+0.0012}$ & 0.74$_{-0.30}^{+0.29}$ & 
0.0036$_{-0.0014}^{+0.0016}$ & 1.08$_{-0.78}^{+0.41}$ & 
    -          &     -        & 
        -       \\
1 & 3.1$\pm$0.6 & 31.2$_{-10.8}^{+43.9}$ & 89.3$_{-13.5}^{+12.2}$ & 0.0515$_{-0.0350}^{+0.2006}$ & 
0.5475$_{-0.3181}^{+0.3822}$ & 3.47$_{-2.11}^{+2.39}$ & 
0.4425$_{-0.3891}^{+0.3170}$ & 1.78$_{-1.19}^{+3.33}$ & 
0.0054$_{-0.0012}^{+0.0013}$ & 0.82$_{-0.63}^{+0.52}$ & 
0.0011$_{-0.0009}^{+0.0014}$ & 0.48$_{-0.38}^{+1.01}$ & 
0.0035$_{-0.0035}^{+0.0331}$ & 0.89$_{-0.83}^{+0.63}$ & 
       -           \\
2 & 6.0$\pm$1.0 & 33.6$_{-10.0}^{+16.2}$ & 134.0$_{-6.8}^{+21.5}$ & 0.1070$_{-0.0649}^{+0.1931}$ & 
0.2973$_{-0.2574}^{+0.1866}$ & 5.16$_{-3.46}^{+0.81}$ & 
0.6921$_{-0.1864}^{+0.2567}$ & 0.47$_{-0.35}^{+1.79}$ & 
0.0101$_{-0.0009}^{+0.0017}$ & 0.47$_{-0.31}^{+0.42}$ & 
0.0003$_{-0.0003}^{+0.0012}$ & 0.1$_{-0.0}^{+0.0}$ & 
0.0002$_{-0.0002}^{+0.0045}$ & 0.1$_{-0.0}^{+0.0}$ & 
       -         \\
3 & 5.2$\pm$0.8 & 26.9$_{-16.9}^{+147.2}$ & 173.8$_{-37.5}^{+343.9}$ & 0.0980$_{-0.0921}^{+0.8132}$ & 
0.1832$_{-0.1675}^{+0.2281}$ & 4.90$_{-4.52}^{+1.07}$ & 
0.7364$_{-0.3726}^{+0.2133}$ & 1.10$_{-0.97}^{+0.95}$ & 
0.0708$_{-0.0529}^{+0.5710}$ & 0.55$_{-0.37}^{+0.42}$ & 
0.0061$_{-0.0061}^{+0.0638}$ & 0.1$_{-0.0}^{+0.0}$ & 
   -    &   -   & 
    -        \\
4 & 6.8$\pm$0.9 & 21.0$_{-5.3}^{+6.2}$ & 104.4$_{-4.5}^{+1.9}$ & 0.0259$_{-0.0108}^{+0.0163}$ & 
0.0866$_{-0.0577}^{+0.0621}$ & 4.37$_{-2.80}^{+1.45}$ & 
0.8956$_{-0.0619}^{+0.0581}$ & 0.28$_{-0.18}^{+0.51}$ & 
0.0154$_{-0.0008}^{+0.0010}$ & 0.32$_{-0.17}^{+0.21}$ & 
      -          &      -        & 
0.0024$_{-0.0023}^{+0.0089}$ & 0.1$_{-0.0}^{+0.0}$ & 
     -          \\
5 & 3.2$\pm$0.6 & 33.2$_{-16.6}^{+39.1}$ & 186.2$_{-33.5}^{+142.1}$ & 0.1327$_{-0.1136}^{+0.3108}$ & 
0.0230$_{-0.0213}^{+0.1883}$ & 2.27$_{-1.28}^{+3.91}$ & 
0.9521$_{-0.1950}^{+0.0218}$ & 2.39$_{-1.95}^{+1.61}$ & 
0.0165$_{-0.0023}^{+0.0043}$ & 0.38$_{-0.25}^{+0.37}$ & 
0.0079$_{-0.0032}^{+0.0049}$ & 0.82$_{-0.61}^{+0.60}$ & 
0.0005$_{-0.0005}^{+0.0053}$ & 0.1$_{-0.0}^{+0.0}$ & 
     -                \\
6 & 4.2$\pm$0.6 & 15.85$_{-5.5}^{+59.8}$ & 174.5$_{-19.7}^{+121.1}$ & 0.0266$_{-0.0193}^{+0.3374}$ & 
0.0448$_{-0.0443}^{+0.1745}$ & 5.67$_{-5.89}^{+0.34}$ & 
0.9216$_{-0.1567}^{+0.0524}$ & 1.09$_{-0.89}^{+0.60}$ & 
0.0193$_{-0.0022}^{+0.0060}$ & 0.76$_{-0.32}^{+0.28}$ & 
   -    &  -  & 
0.0137$_{-0.0101}^{+0.1975}$ & 0.1$_{-0.0}^{+0.0}$ & 
     -               \\
\tableline
\enddata
\end{deluxetable}

\clearpage
\pagestyle{plaintop}

\clearpage
\resizebox{0.8\hsize}{!}{\plotone{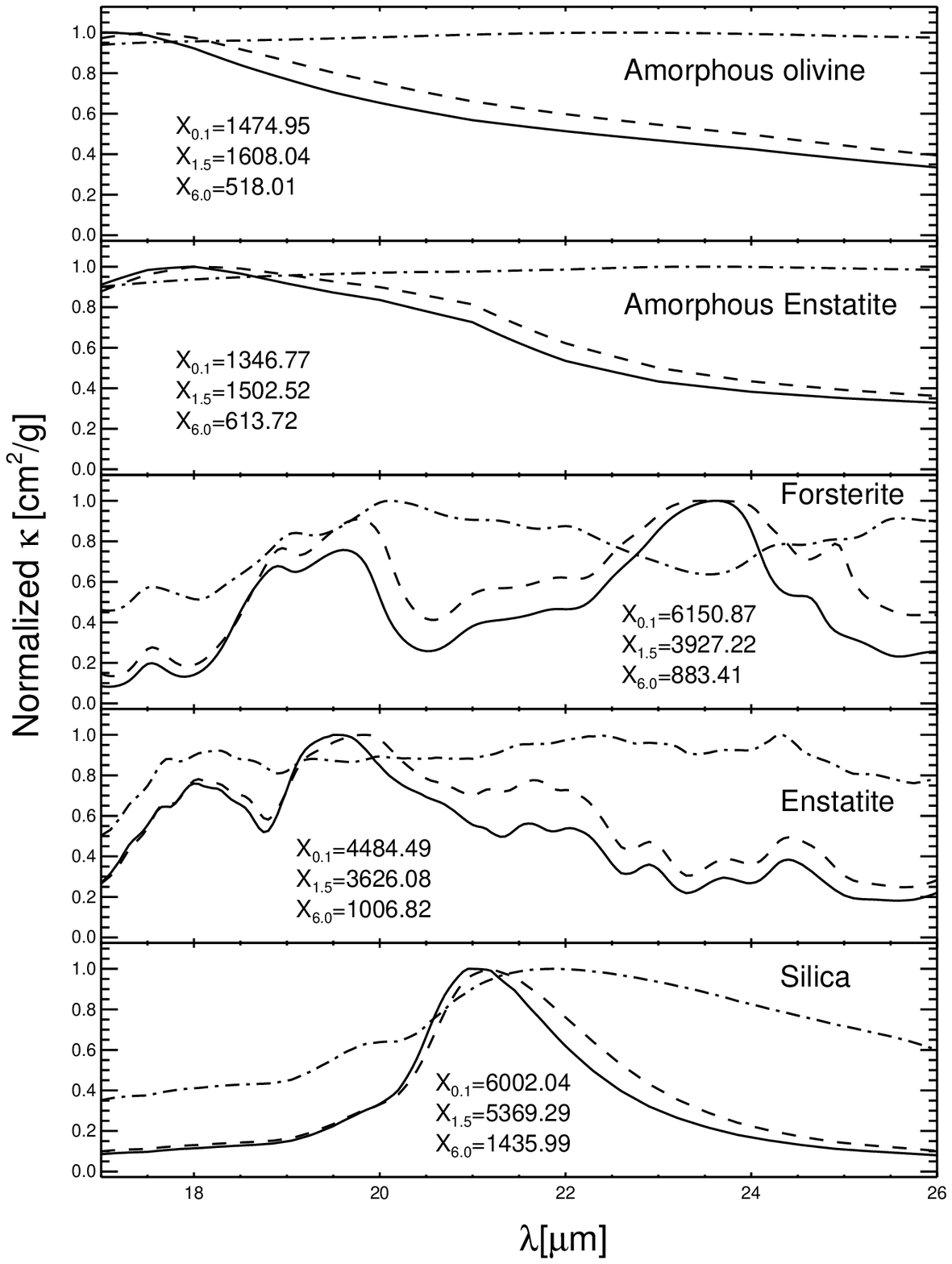}}\\
\centerline{Fig. 3. --- Continued.}
\resizebox{0.8\hsize}{!}{\plotone{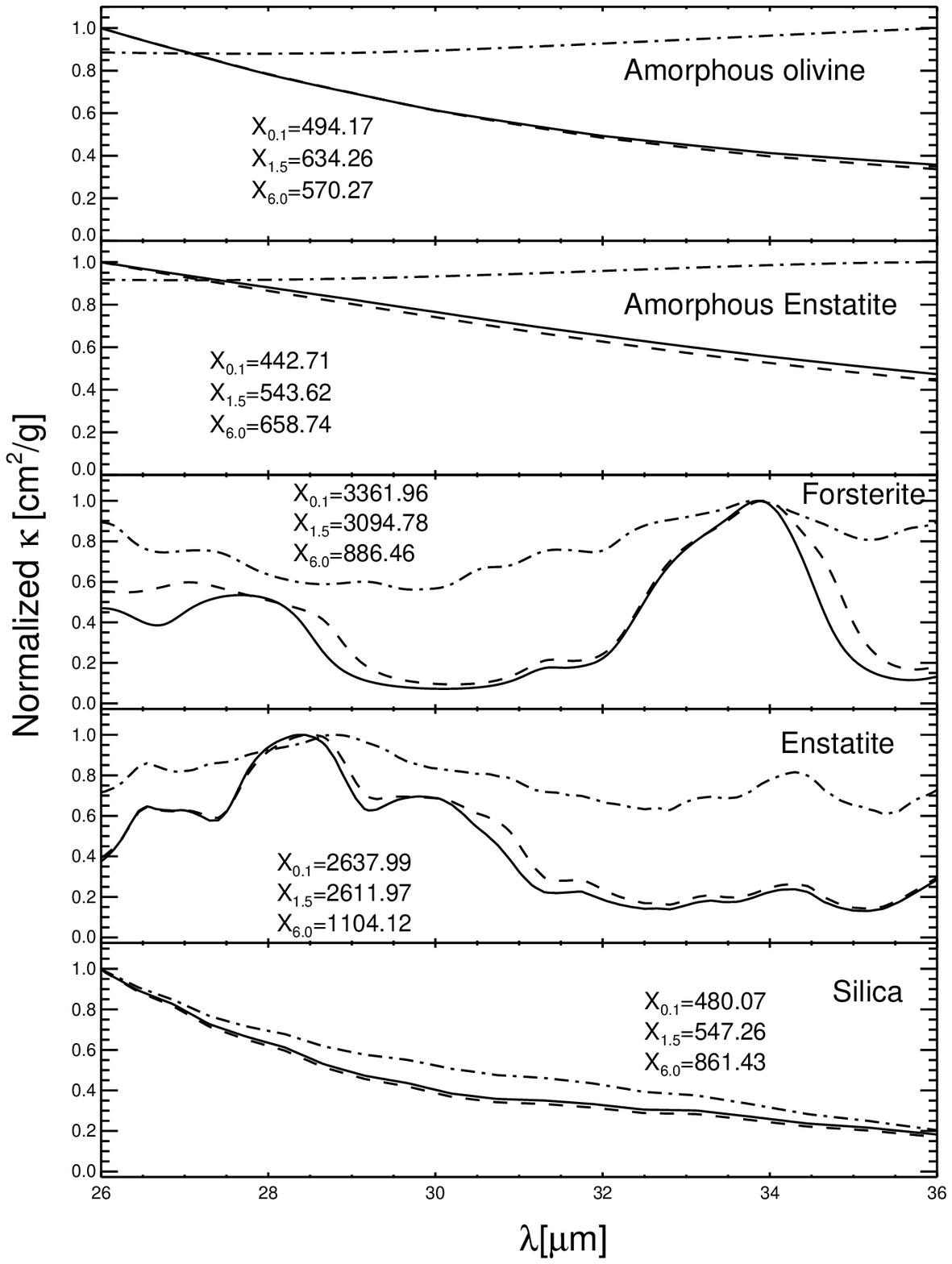}}\\
\centerline{Fig. 3. --- Continued.}

\clearpage

\end{document}